\documentclass[article]{aa} 

\usepackage{graphicx}
\usepackage{natbib}
\bibpunct{(}{)}{;}{a}{}{,} 
\usepackage{txfonts}
\usepackage{booktabs} 
\usepackage{xcolor}
\usepackage{float}
\usepackage{tabularx}
\definecolor{darkgreen}{rgb}{0.0, 0.5, 0.0}
\definecolor{violet}{rgb}{0.5, 0.0, 0.5}
\usepackage[colorlinks,citecolor=blue,linkcolor=red,hyperfootnotes=true,urlcolor=blue]{hyperref}
\usepackage{subfig} 
\usepackage{longtable} 
\usepackage{booktabs}  
\usepackage{array}     
\usepackage{threeparttable}
\usepackage{orcidlink}
\usepackage[normalem]{ulem}
\usepackage[utf8]{inputenc}
\usepackage{siunitx}
\usepackage{multirow}
\usepackage[utf8]{inputenc}
\usepackage{comment}
\usepackage{dblfloatfix} 
\newcommand{\mearth}{M_\oplus}
\newcommand{\rearth}{R_\oplus}
\newcommand{\msun}{M_\odot}
\newcommand{\rsun}{R_\odot}

\providecommand{\nolinenumbers}{}
\begin{document} 

    
    \title{A homogeneous mass campaign of transiting sub-Neptunes around FGK stars}
    \titlerunning{Homogeneous sub-Neptune masses around FGK stars}
    \authorrunning{P. Lamontagne et al.}
    
\author{
Pierrot Lamontagne\inst{1}\and
Rafael Luque\inst{1}\and
Pedro Figueira\inst{1}\and
Carlos Cifuentes\inst{2}\and
Matthew C. Nixon\inst{3}\and
Luis Welbanks\inst{3}\and
Enric Pall\'e\inst{4,5}\and
Ilaria Carleo\inst{6}\and
María Rosa Zapatero Osorio\inst{2}\and
Dawid Jankowski\inst{7}\and
Grzegorz Nowak\inst{7}\and
Natalie M. Batalha\inst{8}\and
Joseph M. Akana Murphy\inst{8}\and
Gaia Lacedelli\inst{4,5}\and
Felipe Murgas\inst{4,5}
}

\institute{
\inst{1}Instituto de Astrof\'isica de Andaluc\'ia (IAA-CSIC), Glorieta de la Astronom\'ia s/n, 18008 Granada, Spain\\
\inst{2}Centro de Astrobiología (CAB), CSIC-INTA, Camino Bajo del Castillo s/n, 28692, Villanueva de la Cañada (Madrid), Spain\\
\inst{3}School of Earth and Space Exploration, Arizona State University, Tempe, AZ 85287, USA\\
\inst{4}Instituto de Astrof\'isica de Canarias (IAC), C/ V\'ia L\'actea s/n, E-38205 La Laguna, Tenerife, Spain\\
\inst{5}Departamento de Astrof\'isica, Universidad de La Laguna (ULL), E-38206 La Laguna, Tenerife, Spain\\
\inst{6}INAF -- Osservatorio Astrofisico di Torino, Via Osservatorio 20, I-10025 Pino Torinese, Italy\\
\inst{7}Institute of Astronomy, Faculty of Physics, Astronomy and Informatics, Nicolaus Copernicus University, Grudziądzka 5, 87-100 Toruń, Poland\\
\inst{8}Department of Astronomy and Astrophysics, University of California, Santa Cruz, CA 95064, USA\\
}

\date{Received 29 July 2026 / Accepted 16 November 2026}

\abstract{
Sub-Neptunes are among the most common planets in our Galaxy, but are still poorly understood, with fundamental questions about their origins and compositions remaining unanswered. Addressing these questions requires precise bulk densities, but current literature values are reported across studies employing different radial velocity (RV) extraction methods, stellar activity treatments, and statistical frameworks. These heterogeneities introduce systematic biases in the derived parameters that can lead to spurious conclusions in population-level analyses. We conducted a systematic and homogeneous mass-measurement campaign for 12 sub-Neptune systems observed by the Kepler mission and its K2 extension, selected for their absent or poorly constrained mass measurements ($\sigma_{M_p}/M_p > 25\%$). Using new high-resolution spectroscopy from HIRES, HARPS-N, and ESPRESSO, as well as archival data, we applied a unified hierarchical Bayesian framework with homogeneous data curation. For HARPS-N and ESPRESSO echelle spectra, we applied four independent extraction pipelines (CCF, SERVAL, LBL, sBART) and combined the resulting posteriors via leave-one-out cross-validation stacking, so that no mass would hinge on a single pipeline. In addition, a preferred noise model for each target was selected through a uniform Bayesian procedure. Thus, differences among the derived masses could not stem from specific choices made in the analysis. We obtained mass determinations for four transiting planets, Kepler-21\,b ($M_p = 5.66^{+0.65}_{-0.96}\,M_\oplus$, $\rho_p = 7.61^{+0.96}_{-1.28}\,\mathrm{g\,cm^{-3}}$), K2-66\,b ($17.9 \pm 3.2\,M_\oplus$, $4.69 \pm 1.31\,\mathrm{g\,cm^{-3}}$), Kepler-538\,b ($10.8^{+2.4}_{-2.8}\,M_\oplus$, $5.47 \pm 1.22\,\mathrm{g\,cm^{-3}}$), and, more marginally, K2-182\,b ($20.1^{+4.5}_{-7.8}\,M_\oplus$, $5.33^{+1.29}_{-2.06}\,\mathrm{g\,cm^{-3}}$). We also provide updated masses for the non-transiting giants Kepler-65\,e and Kepler-129\,d, and report upper limits for the remaining 15 signals, including the first constraints for K2-244 and K2-269. No new non-transiting companion was detected in any system, which is consistent with the fact that massive planets on short-to-intermediate periods do not dominate the architecture of these sub-Neptune systems. Two planets lie $\approx2\sigma$ from their published masses: Kepler-126\,d, with an RV upper limit of $17\,M_\oplus$ against a minimum mass of $55 \pm 23\,M_\oplus$, and Kepler-145\,b, with $21\,M_\oplus$ against a transit timing variation mass of $34^{+25}_{-11}\,M_\oplus$. Across the multi-pipeline targets, the semi-amplitudes of the four reductions span a median of $0.75\sigma$ of the stacked posterior and up to $2.7\sigma$, which is a systematic that a single-pipeline analysis would ignore.
}


\maketitle
\nolinenumbers

\section{Introduction}\label{sec:intro}

Sub-Neptunes ($1.5\,R_{\oplus} < R_{\mathrm{p}} < 4\,R_{\oplus}$) are among the most common outcome of planet formation in the Galaxy, yet their physical nature remains ambiguous. Transit surveys (primarily Kepler) have revealed a bimodality in the planetary radius distribution of close-in planets orbiting solar-type stars, with a dearth of planets between $1.5$ and $2.0\,R_{\oplus}$, in the so-called radius valley \citep{Fulton2017}. This valley is commonly attributed to atmospheric mass loss via photoevaporation \citep{Owen2013} or core-powered mass loss \citep{Gupta2019}, though \citet{Raymond2018, Venturini2020, Burn2024, Nielsen2025} and others showed that a valley-like feature can also emerge during planet formation itself, with these two mass-loss channels subsequently reshaping the resulting population. This feature is often interpreted as the boundary between predominantly rocky super-Earths and sub-Neptune planets with non-negligible volatile mass. Indeed, interior models indicate that planets larger than $\sim1.6\,R_{\oplus}$ are unlikely to be purely rocky \citep{Rogers2015}. Yet radius alone is an incomplete diagnostic in this radius regime: both a rocky core with a modest H/He envelope and an ice--rock mixture with little primordial gas can be characterised by the same radius \citep[e.g.][]{Rogers2010}. Even obtaining the mass of an individual exoplanet to measure the bulk density is not enough to differentiate between these two composition scenarios. A single density measurement is not sufficient, as a homogeneously measured bulk density sample would be required to test against the predictions of formation models \citep{Burn2024b}. Such tests have been carried out on compiled samples of M-dwarf planets \citep{Luque2022,Parc2024}, but doubts remain as to whether density differences reflect heterogeneous analyses, rather than the nature of the planets.

Around M dwarfs, two distinct density populations emerge, separating predominantly rocky planets from bodies consistent with half-rock-half-ice compositions \citep{Luque2022, Parviainen2023, Parc2024, Lacedelli2026}. Theoretical models demonstrate that such volatile-rich planets can reproduce the observed bimodality without invoking atmospheric escape as the sole sculpting mechanism; for instance, through migration-driven assembly and steam-world evolution \citep{Venturini2020,Venturini2024,Burn2024,Chakrabarty2026}. Yet the interpretation remains contested since a rocky core retaining a thin H/He envelope produces bulk densities indistinguishable from those of a water-rich body at the precision currently achievable, leaving the gas-dwarf scenario viable for the same observations \citep[e.g.][]{Dorn2015,Zeng2019,Dorn2021,Rogers2023}. Moreover, the transition between super-Earths and sub-Neptunes appears to vary with stellar mass both observationally \citep{Parviainen2023,Parc2024} and theoretically \citep[e.g.][]{Burn2021,Venturini2024}, hinting that host-star environment plays a central role in shaping planetary composition. For low-mass stars, the density and radius gaps seem to close up because the combination of prolonged high extreme-ultraviolet (XUV) irradiation, lower disk mass budgets, and the efficient migration of planets formed beyond the snowline produces super-Earth and sub-Neptune populations that appear more alike in mass and radius \citep{Cloutier2020, Luque2022, Parc2024, Venturini2024, Gillis2026}. 

In contrast, planets around higher-mass FGK stars effectively conserve initial formation pathways better, preserving the distinct signatures of accretion and migration \citep{Ho2024}. Establishing whether FGK-hosted sub-Neptunes are predominantly gas dwarfs \citep{Rogers2023} or water-rich bodies \citep{Burn2024} is therefore a direct test of formation pathways. This density-based test is particularly useful because atmospheric characterisation (the only other way to break the degeneracy) can be ambiguous for M-dwarf exoplanets observed with the James Webb Space Telescope (JWST). Cloud-shrouded, hazy, or high-metallicity hydrogen envelopes on gas dwarfs can effectively mimic the transmission spectra of true water-rich worlds \citep{Roy2023, Piaulet2024, Roy2025}. A census of planets around FGK stars would provide the community with a large, unbiased sample on which to test planet formation models that predict different underlying planetary compositions.

Drawing on published RV masses for sub-Neptunes from a number of differing reduction pipelines, stellar-activity treatments, and statistical frameworks can often inject poorly quantified systematics into the mass--radius diagram. Thus, apparent compositional branches or extreme outliers may reflect analysis choices, rather than the astrophysical reality \citep{Burt2018,Osborne2024,AkanaMurphy2024}. In this study, we establish a precedent for homogeneous RV analyses by presenting a mass-measurement campaign for 12 sub-Neptune systems orbiting bright ($V < 12.5$\,mag) FGK dwarfs, selected precisely because their published masses are either unconstrained or yield `outlier' bulk compositions. Here, we use the term `homogeneous' to refer to the analysis specifically. The number, cadence, and precision of the RVs differ between systems, but the data vetting, noise-model selection, priors, and stacking remain the same for all twelve. All targets are analysed within a unified hierarchical Bayesian framework, with consistent RV extraction across four independent reduction pipelines. Crucially, rather than arbitrarily selecting a single reduction, we combine the resulting posteriors using leave-one-out cross-validation stacking \citep{Yao2018}, a principled approach that weights each pipeline according to its predictive performance on the data itself. The result is a set of homogeneously derived masses that can be placed in the mass--radius diagram with confidence.

\section{Sample selection}\label{sec:sample}

\begin{figure}
\centering
\includegraphics[width=0.90\columnwidth]{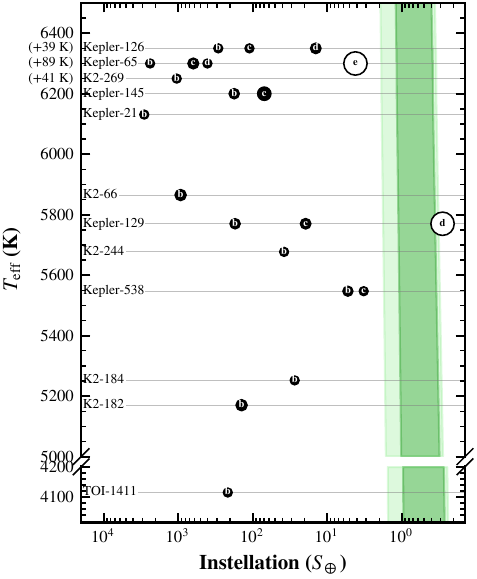}
\caption{System architecture of the sample: 21 planets, including two non-transiting giants (radii unmeasured, drawn at one Jupiter radius), around hosts with $T_\mathrm{eff}$ between 4100 and 6300\,K. Habitable-zone boundaries are from \citet{Kopparapu2013}.}
\label{fig:system_architecture}
\end{figure}

We used the NASA Exoplanet Archive to identify sub-Neptunes whose existing mass determinations are either absent or insufficiently precise to constrain the planet composition via its location on the mass-radius (M--R) diagram. Specifically, we required a fractional mass precision of $\sigma_{M_p}/M_p$ that was worse than 25\%, so that each target could benefit from a renewed RV analysis. The planet size was restricted to $R_p < 4\,R_\oplus$ and orbital periods to $P > 1$\,d. To ensure that the RV signal is detectable with current instrumentation, we required an expected semi-amplitude of $K > 1.5$\,m\,s$^{-1}$. The expected semi-amplitudes for planets with no published mass were estimated from the probabilistic M--R relation of \citet{Parviainen2023} (\textsc{spright}, \texttt{tepcat\_fgk\_z19} model trained on FGK hosts). Furthermore, we required a fractional radius precision of better than 15\% to prevent the density error budget from being dominated by radius uncertainty. Host stars were restricted to the FGK main sequence as well as evolved stars with $T_{\rm eff} > 4000$\,K and $V < 12.5$\,mag, which gives sufficient photon flux for precise RV work at $\sim$1\,m\,s$^{-1}$ with 4--8\,m class telescopes. Within these constraints, we prioritised systems occupying sparsely sampled regions of the mass--radius diagram, including candidate superdense sub-Neptunes and super-Mercury analogues, where new mass measurements carry the greatest discriminating power between competing interior models. The resulting sample comprises 12 planetary systems containing 21 total planets whose architectures are shown in Fig.~\ref{fig:system_architecture}. Table~\ref{tab:sample} summarises the basic stellar properties of our sample. For each system, we adopted the single reference providing the most internally consistent stellar characterisation, prioritising dedicated spectroscopic analyses or homogeneous stellar studies when available and otherwise referring to the original planet discovery paper. To ensure that the stellar parameters adopted for the sample followed a homogeneous approach, we independently derived $T_\mathrm{eff}$, $R_\star$, and $M_\star$ for every target via broadband spectral energy distribution (SED) fitting with \texttt{ariadne} \citep{Vines2022} and compared the results against the adopted literature values in Fig.~\ref{fig:ariadne_comparison}. The two sets of parameters are in agreement: the median offset over all $T_\mathrm{eff}$, $R_\star$, and $M_\star$ comparisons is $0.64\sigma$, while the largest is $2.3\sigma$, for the isochrone mass of TOI-1411, whose $T_\mathrm{eff}$ and $R_\star$ also differ by $1.7\sigma$ and $2.1\sigma$. TOI-1411 is the coolest and smallest star of the sample ($T_\mathrm{eff}\approx4100$\,K, $R_\star\approx0.61\,R_\odot$), approaching the regime where the SED fits of \texttt{ariadne} show increasing systematic offsets \citep{Vines2022}. This overall agreement shows that the adopted literature parameters are consistent with an independent and homogeneous determination. We therefore retained them as the sample's reference. The \texttt{ariadne} values would not improve on them, since its isochrone masses are less precise than the literature ones for 11 of the 12 targets (by a median factor of 2.3). 

\begin{table*}
\setlength{\tabcolsep}{4pt}
\caption{Stellar properties of the target sample.}
\label{tab:sample}
\centering
\resizebox{\textwidth}{!}{%
\begin{tabular}{l c c c c c c c c c}
\hline\hline
Star & $N_{\rm pl}$ & $V$ (mag) & Sp. Type & $M_\star$ ($\msun$) & $R_\star$ ($\rsun$) & $T_{\rm eff}$ (K) & [Fe/H] & $v\sin i$ (km\,s$^{-1}$) & Ref. \\
\hline
K2-182 & 1 & $11.833\pm0.069$ & G9V  & $0.865^{+0.016}_{-0.018}$ & $0.793^{+0.013}_{-0.011}$ & $5170 \pm 100$ & $0.12\pm0.06$ & $<2$ & Aka21 \\
K2-184 & 1 & $10.345\pm0.009$ & G8V  & $0.819^{+0.028}_{-0.030}$ & $0.764^{+0.033}_{-0.025}$ & $5252 \pm 50$ & $-0.22\pm0.08$ & $1.51$\tablefootmark{b} & May18 \\
K2-244 & 1 & $11.780\pm0.024$ & G3V  & $0.86\pm0.02$ & $0.92^{+0.02}_{-0.01}$ & $5677^{+38}_{-39}$ & $-0.29\pm0.03$ & $3.69\pm0.24$ & Liv18 \\
K2-269 & 1 & $11.784\pm0.028$ & F7V  & $1.16\pm0.04$ & $1.45\pm0.06$ & $6209 \pm 88$ & $-0.080\pm0.060$ & $6.1$\tablefootmark{b} & Liv18b \\
K2-66  & 1 & $11.710\pm0.026$ & G1IV & $1.06\pm0.07$ & $1.79\pm0.10$ & $5865 \pm 100$ & $-0.050\pm0.044$ & $4.0\pm1.0$ & How25 \\
Kepler-126 & 3 & $10.505\pm0.006$ & F7IV & $1.180^{+0.045}_{-0.058}$ & $1.350^{+0.031}_{-0.036}$ & $6311^{+81}_{-75}$ & $-0.070^{+0.078}_{-0.084}$ & $8.7\pm1.0$\tablefootmark{b} & Mor16 \\
Kepler-129 & 3 & $11.809\pm0.022$ & G4IV & $1.178^{+0.021}_{-0.030}$ & $1.653^{+0.009}_{-0.012}$ & $5770 \pm 83$ & $0.29\pm0.10$ & $2.13\pm0.10$ & Zha21 \\
Kepler-145 & 2 & $11.977\pm0.034$ & F4V  & $1.370^{+0.042}_{-0.060}$ & $1.890^{+0.047}_{-0.056}$ & $6200^{+100}_{-72}$ & $0.100^{+0.080}_{-0.067}$ & $6.7\pm1.0$\tablefootmark{b} & Mor16 \\
Kepler-21  & 1 & $8.25\pm0.03$ & F6IV & $1.340\pm0.010$ & $1.860\pm0.020$ & $6131 \pm 44$ & $-0.15\pm0.06$ & $7.75\pm1.00$ & How12 \\
Kepler-538 & 2\tablefootmark{a} & $11.270\pm0.018$ & G5V & $0.892^{+0.051}_{-0.035}$ & $0.8717^{+0.0064}_{-0.0061}$ & $5547 \pm 50$ & $-0.030\pm0.080$ & $1.10\pm0.50$ & May19 \\
Kepler-65  & 4 & $10.922\pm0.008$ & F6IV & $1.25\pm0.06$ & $1.41\pm0.03$ & $6211 \pm 66$ & $0.17\pm0.06$ & $11.9\pm0.5$ & Cha13 \\
TOI-1411   & 1 & $10.51\pm0.03$ & K7V & $0.6045^{+0.0148}_{-0.0129}$ & $0.6140^{+0.0185}_{-0.0169}$ & $4115^{+42}_{-45}$ & $0.012^{+0.071}_{-0.075}$ & $2.00\pm0.50$\tablefootmark{b} & Mac23 \\
\hline
\end{tabular}%
}
\tablefoot{
Reference abbreviations:
Aka21 = \citet{AkanaMurphy2021};
May18 = \citet{Mayo2018};
Liv18 = \citet{Livingston2018};
Liv18b = \citet{Livingston2018b};
How25 = \citet{Howard2025};
Mor16 = \citet{Morton2016};
Zha21 = \citet{Zhang2021};
How12 = \citet{Howell2012};
May19 = \citet{Mayo2019};
Cha13 = \citet{Chaplin2013};
Mac23 = \citet{MacDougall2023}.
Spectral types were revised using a \textit{Gaia}-based reassessment.
\tablefootmark{a}Includes the unconfirmed candidate Kepler-538\,c (KOI-365.02).
\tablefootmark{b}$v\sin i$ from \citet{Abdurrouf2022} for K2-184, \citet{Petigura2017} for K2-269, \citet{Petigura2022} for Kepler-126 and Kepler-145, and \citet{Giacalone2022} for TOI-1411.
}
\end{table*}

\subsection*{Prior work on the target systems}

The literature masses, $M_p^{\rm lit}$, which we used for the comparison presented in Sect.~\ref{sec:masses}, are listed in Table~\ref{tab:masses}, with their source. Below, we summarise, for each target, the discovery and validation of its planets and the published mass measurements.

K2-182: K2-182\,b was identified as a K2 Campaign 5 candidate \citep{Pope2016,Petigura2017} and validated by \citet{Mayo2018}. \citet{AkanaMurphy2021} measured $20\pm5\,\mearth$ from 12 RVs of the Keck High Resolution Echelle Spectrometer (HIRES), which indicates the planet is a superdense sub-Neptune. \citet{Thygesen2023} refit these RVs with Transiting Exoplanet Survey Satellite (TESS) transits and found $21.0^{+5.4}_{-5.1}\,\mearth$, which is the mass we used for our comparison.

K2-184, K2-244, and K2-269: K2-184\,b, a Campaign 5 candidate \citep{Barros2016,Pope2016}, was validated by \citet{Mayo2018}, and its transit timing variations (TTVs) suggest a non-transiting companion \citep{CastroGonzlez2021}. \citet{Hobson2026} measured $10.9\pm4.7\,\mearth$ ($K=4.4^{+1.5}_{-1.6}\,\mathrm{m\,s^{-1}}$) for K2-184\,b from ESPRESSO RVs and K2 photometry. K2-244\,b and K2-269\,b were validated by \citet{Livingston2018} and \citet{Livingston2018b}, and neither has a published mass.

K2-66: K2-66\,b was identified as a K2 candidate \citep{Vanderburg2016} and validated by \citet{Crossfield2016}. \citet{Sinukoff2017} measured $21.3\pm3.6\,\mearth$ from 38 Keck/HIRES RVs, and \citet{Howard2025} revised it to $16\pm4\,\mearth$ from 44 HIRES RVs that include those 38, which places the planet in the photoevaporation desert \citep{Lundkvist2016}. We used the latter in our comparison.

Kepler-126: The three transiting planets were validated by \citet{Rowe2014} and show TTVs \citep{Holczer2016}. The only published masses are minimum masses from 58 Keck/HIRES RVs \citep{Weiss2023}: $4\pm7$, $1\pm7$, and $55\pm23\,\mearth$ for b, c, and d, respectively.

Kepler-129: Planets b and c were validated by \citet{Rowe2014}. \citet{Zhang2021} found the giant Kepler-129\,d in Keck/HIRES RVs and measured $43^{+13}_{-12}\,\mearth$ for c and an upper limit of $20\,\mearth$ for b. With a longer HIRES baseline, \citet{Weiss2023} obtained $P_d = 1936\pm25$\,d and $M_d\sin i = 2015\pm43\,\mearth$, and minimum masses of $10\pm6$ and $25\pm9\,\mearth$ for b and c, which we used for our comparison. 

Kepler-145: \citet{Xie2014} confirmed both planets from their TTVs, with masses of $37\pm12$ and $79\pm16\,\mearth$. A photodynamical fit of the Kepler photometry gives $34^{+25}_{-11}$ and $25.1^{+9.4}_{-6.0}\,\mearth$ for b and c \citep{Ofir2025}. No RV mass has been published from the 26 Keck/HIRES RVs of \citet{Weiss2023}.

Kepler-21: \citet{Howell2012} validated the planet, placed an upper limit of $10.4\,\mearth$ on its mass from Keck/HIRES RVs and characterised the host with asteroseismology. \citet{LopezMorales2016} measured $5.1\pm1.7\,\mearth$ from High Accuracy Radial velocity Planet Searcher for the Northern hemisphere (HARPS-N) and HIRES RVs, and \citet{Bonomo2023} $7.5\pm1.3\,\mearth$ from a homogeneous reanalysis of the HARPS-N Guaranteed Time Observations (GTO) RVs. \citet{BeardRobertson2024} added NEID RVs, obtained $6.68\pm1.4\,\mearth$ from a joint fit of the Kepler and TESS photometry and the HIRES, HARPS-N, and NEID RVs, and attributed a long-term RV drift to a candidate outer planet. We used their most recent mass result for our comparison. 

Kepler-538: Kepler-538\,b was validated by \citet{Morton2016}. \citet{Mayo2019} measured $10.6^{+2.5}_{-2.4}\,\mearth$ from HARPS-N and HIRES RVs, with a Gaussian process (GP) for stellar activity, and \citet{Bonomo2023} $12.9\pm2.9\,\mearth$ from the same reanalysis; our comparison employed the latter. The candidate Kepler-538\,c (KOI-365.02), a weak 117.8-d transit signal listed in early \textit{Kepler} catalogues, but absent from the final Data Release 25 (DR25) catalogue \citep{Mayo2019}, has no published mass.

Kepler-65: \citet{Chaplin2013} confirmed the three transiting planets and characterised the host with asteroseismology. \citet{Mills2019} found the non-transiting giant Kepler-65\,e in Keck/HIRES RVs and derived the masses of b, c, and d from a photodynamical fit of the Kepler photometry and the RVs. We compared with these and with the $M_e\sin i = 207.5^{+17.8}_{-17.5}\,\mearth$ of Kepler-65\,e. 

TOI-1411: TOI-1411\,b was validated by \citet{Giacalone2022}, with stellar parameters from \citet{MacDougall2023}. \citet{Polanski2024} measured $2.0^{+1.2}_{-1.1}\,\mearth$ from Keck/HIRES RVs of the TESS-Keck Survey. We used this mass in our comparison.

\section{Observations and data reduction} \label{sec:observations}

For the 12 systems described above, we used RVs from five instruments, ordered here by number of measurements: HIRES, HARPS-N, the Echelle SPectrograph for Rocky Exoplanets and Stable Spectroscopic Observations (ESPRESSO), the Automated Planet Finder (APF), and NEID. Table~\ref{tab:observations} lists the counts per target and instrument, split into new data (not previously published and not from a public archive) and archival data. The photometric data are described in Sect.~\ref{sec:photometry_analysis}.

\begin{table}[t]
\caption{RV observations per target and instrument after quality vetting (Sect.~\ref{sec:vetting}), as new\,+\,archival.}
\label{tab:observations}
\centering
\footnotesize
\setlength{\tabcolsep}{3pt}
\begin{tabular}{lrrrrrr}
\toprule
System & HIRES & HARPS-N & ESPRESSO & APF & NEID & Total \\
\midrule
K2-182 & 2+12 & 0+4 & -- & -- & -- & 18 \\
K2-184 & -- & 36+0 & 0+45 & -- & -- & 81 \\
K2-244 & -- & 46+0 & -- & -- & -- & 46 \\
K2-269 & -- & 21+0 & 36+0 & -- & -- & 57 \\
K2-66 & 12+44 & -- & -- & -- & -- & 56 \\
Kepler-21 & 7+66 & 0+97 & -- & 59+0 & 0+20 & 249 \\
Kepler-65 & 0+79 & 45+0 & -- & -- & -- & 124 \\
Kepler-126 & 0+58 & 64+0 & -- & -- & -- & 122 \\
Kepler-129 & 0+55 & -- & -- & -- & -- & 55 \\
Kepler-145 & 0+26 & -- & -- & -- & -- & 26 \\
Kepler-538 & 0+46 & 0+83 & -- & -- & -- & 129 \\
TOI-1411 & 24+45 & -- & -- & -- & 0+7 & 76 \\
\midrule
Total & 45+431 & 212+184 & 36+45 & 59+0 & 0+27 & 1039 \\
\bottomrule
\end{tabular}
\end{table}

\subsection{Keck/HIRES}

We used HIRES, a cross-dispersed echelle spectrograph mounted on the Keck-I 10\,m telescope on Maunakea, Hawaii \citep{Vogt1994}. HIRES provides a resolving power of $R\approx67{,}000$ and achieves a precision of a few m\,s$^{-1}$ using an iodine absorption cell for simultaneous wavelength calibration. We used 476 HIRES RVs of nine targets. Of these, 431 were published \citep{Weiss2023, Howard2025, AkanaMurphy2021, Polanski2024, BeardRobertson2024}, and 45 are previously unpublished epochs from the California Planet Search (CPS).\footnote{Includes programme 2023A\_U156.} All were taken from the CPS database in its current iodine reduction, so published and unpublished RVs share one reduction. This reduction forward-models each observed spectrum as a Doppler-shifted, deconvolved iodine-free stellar template multiplied by the iodine-cell transmission function and convolved with the spectrometer instrumental profile \citep{Butler1996, Howard2010}.

\subsection{TNG/HARPS-N}

HARPS-N is a fibre-fed echelle spectrograph mounted on the 3.6\,m Telescopio Nazionale Galileo (TNG) at the Roque de los Muchachos Observatory, La Palma, Spain, with a resolving power of $R \simeq 115{,}000$ \citep{Cosentino2012}. We obtained new spectra of five targets (K2-184, K2-244, K2-269, Kepler-65, and Kepler-126)\footnote{Programmes CAT23A\_52, CAT23B\_74, CAT24B\_20, and CAT25A\_76.} and used archival spectra of Kepler-21 and Kepler-538 from the HARPS-N GTO\footnote{Programme CAT15B\_79 and the HARPS-N GTO.} and four archival spectra of K2-182. New and archival spectra were reduced with the same four pipelines and versions (Sect.~\ref{rv_extraction_methods}).

\subsection{VLT/ESPRESSO}

ESPRESSO is a fibre-fed echelle spectrograph installed on the European Southern Observatory (ESO) Very Large Telescope (VLT) at Paranal Observatory \citep{Pepe2021}. All spectra were taken in the standard high-resolution 1-UT (one unit telescope) mode, with a resolving power of $R \simeq 140{,}000$. We obtained 36 new spectra of K2-269\footnote{Programme 112.25F2.001.} and used 45 archival spectra of K2-184, also analysed by \citet{Hobson2026}\footnote{Programmes 0102.C-0456, 1102.C-0744, 1102.C-0958, 1104.C-0350, and 106.21M2.003.}. The latter straddle the 2019 intervention on the fibre link, so we split them into two datasets, ESPRESSO18 (38 spectra) and ESPRESSO19 (7 spectra), fitted with separate RV offsets.

\subsection{APF/Levy}

The APF is a dedicated 2.4\,m telescope at Lick Observatory, California, feeding the Levy spectrometer, a high-resolution echelle spectrograph ($R \simeq 110{,}000$ with the standard 1\arcsec\ slit) that uses an iodine absorption cell for precise Doppler work \citep{Vogt2014}. We used 59 previously unpublished APF RVs of Kepler-21, obtained under programmes 2019A (PI: P.~Robertson) and 2023A (PI: J.~Luhn) at Lick Observatory and taken from the CPS iodine reduction as for HIRES.

\subsection{WIYN/NEID}

NEID is a fibre-fed high-resolution echelle spectrograph installed on the 3.5\,m WIYN Telescope at Kitt Peak National Observatory \citep{Schwab2016,Robertson2019}. It provides a resolving power of $R \simeq 110{,}000$ and uses a laser frequency comb for wavelength calibration. We used 27 archival NEID RVs. The 20 RVs of Kepler-21 are the published values of \citet{BeardRobertson2024}, and the 7 RVs of TOI-1411 are standard cross-correlation function (CCF) RVs from the NEID Data Reduction Pipeline (DRP), retrieved from the NEID archive (programme 2022A-361743).

\subsection{RV extraction}
\label{rv_extraction_methods}
For targets observed with HARPS-N and ESPRESSO, we used RVs derived using different RV extraction pipelines (cross-correlation function, CCF; SpEctrum Radial Velocity AnaLyser, SERVAL; line-by-line, LBL; semi-Bayesian approach for RVs with template matching, sBART). This allowed us to account for pipeline-dependent systematics in a statistically consistent manner. HIRES, APF, and NEID RVs enter as single fixed datasets common to every reduction, so reductions differ only in their HARPS-N and ESPRESSO RVs. This is because iodine RVs (HIRES, APF) can only be obtained by forward modelling and because the pipelines do not yet support NEID.

\subsubsection{CCF}
The CCF method is a standard technique for RV extraction in high-resolution spectroscopy. It consists of correlating the observed stellar spectrum with a numerical mask composed of binary transmission zones (1/0) marking the rest-frame positions of known stellar absorption lines \citep{Baranne1996}. The mask is shifted through velocity space, with the convolution producing an average line profile. The stellar RV is obtained by fitting a Gaussian to the CCF profile \citep{Pepe2002}. To optimise Doppler information, modern implementations employ weighted masks where each line is assigned a weight proportional to its relative depth \citep{Bouchy2001}. The method is sensitive to telluric contamination, which the Data Reduction Software (DRS) excludes entirely from the CCF mask rather than correcting for, discarding up to $\sim$27\% of the spectral range \citep{Allart2022}. In this work, RVs were extracted using v3.3.12 for HARPS-N and v3.3.10 for ESPRESSO of their respective DRS. Each target was correlated with the closest available discrete spectral-type mask to its classification (Table~\ref{tab:sample}).

\subsubsection{SERVAL}
\texttt{SERVAL} is a template-matching algorithm \citep{AngladaEscud2012} that relies on forward modelling in pixel space, decomposing each observation into three components: a high signal-to-noise (S/N) stellar template, $F(\lambda)$; RV shifts, $v_n$, and multiplicative background polynomials, $p(\lambda)$, which account for flux variations arising from effects such as wavelength-dependent fibre coupling losses \citep{Zechmeister2017}. The stellar template is constructed directly from the observations, co-adding all available spectra using weighted least-squares regression with cubic B-splines. Each echelle order is fitted independently by modelling the observed flux against the template defined on a uniform grid in logarithmic wavelength space. Exposures are corrected for telluric contamination and sky emission by masking spectral regions with telluric features deeper than 5\% and known sky lines. The RVs were extracted using the public repository\footnote{\url{https://github.com/mzechmeister/serval}}, using commit \texttt{58e5419}. The same processing configuration was applied to the spectra, using the instrument-specific \texttt{SERVAL} modes for each spectrograph. We kept the default spline oversampling factor \texttt{ofac}\,$=1$ for the template co-addition, since no star in the sample rotates fast enough (Table~\ref{tab:sample}) for its broadened line profiles to require finer knot spacing.

\subsubsection{LBL}
The LBL framework generalises the optimal-weight formalism of \citet{Bouchy2001} by operating at the level of individual spectral features. The spectrum is partitioned into thousands of narrow domains defined by the regions between consecutive local maxima in a high-S/N stellar template. Only extrema at which the template derivative is significant at the $3\sigma$ level on both sides are retained, so that noise in the template does not generate spurious line boundaries. Each domain, effectively corresponding to an individual absorption line, is independently matched to the template to estimate its Doppler shift, yielding a distribution of per-line velocities \citep{Artigau2022}. The final RV is obtained by combining the individual line shifts with a mixture model composed of a nominal population used to create the overall RV measurement and an outlier population that accounts for lines affected by systematic errors. This approach enables automatic identification and down-weighting of lines affected by detector defects, cosmic rays, or residual telluric contamination, without discarding entire wavelength intervals. RVs were extracted with the public repository\footnote{\url{https://github.com/njcuk9999/lbl}}, \texttt{LBL} version \texttt{v0.68.001}. Tellurics were removed beforehand rather than concurrently, in a dedicated two-pass \texttt{telluclean} step that models the atmospheric transmission against Transmissions of the AtmosPhere for AStronomical data (TAPAS; \citealt{Bertaux2014}) ahead of template construction and RV extraction.

\subsubsection{sBART}

\texttt{sBART} recasts classical template matching within a semi-Bayesian framework to provide a probabilistic estimate of RVs and their uncertainties \citep{Silva2022}. In contrast to traditional pipelines that treat echelle orders independently, \texttt{sBART} adopts a global parametrisation in which a single Doppler shift describes all spectral orders simultaneously. The method analytically marginalises over continuum nuisance parameters (e.g. slope and intercept terms) and explicitly propagates uncertainties from the stellar template construction into the final RV estimate. Exposures are corrected for telluric contamination by masking wavelength regions with telluric features deeper than 1\% in a TAPAS transmittance profile. RVs were extracted with the public repository\footnote{\url{https://github.com/iastro-pt/sBART}}, \texttt{sBART} release \texttt{v1.0.2}, adopting a uniform velocity sampling of \(0.1\,\mathrm{m\,s^{-1}}\) and a symmetric search interval of \(\pm 200\,\mathrm{m\,s^{-1}}\).

\subsection{Quality vetting}\label{sec:vetting}

We applied one rule to new and archival RVs alike: manual flags, a $5\sigma$ clip in each reduction, and a common-epoch cut across reductions. The manual flags are two K2-269 epochs with low S/N (13.8 and 9.9), which return unphysical RVs in all pipelines (the first is $\approx+220\,\mathrm{m\,s^{-1}}$ off). The common-epoch cut removes an epoch from all reductions when any pipeline's own quality control rejects it, which keeps the reductions on identical spectra and is conservative. All 17 epochs removed are HARPS-N: the two flagged, four clipped (two in Kepler-538, one each in K2-184 and K2-244), and eleven cut as common-epoch rejects (seven in Kepler-65, two in Kepler-126, one each in Kepler-21 and Kepler-538).

\section{Photometric analyses}\label{sec:photometry_analysis}

Photometry from \textit{Kepler}, K2, and TESS was used to infer prior distributions for stellar activity parameters in the RV modelling, for the targets whose adopted noise model includes a GP (Sect.~\ref{sec:modelling}). Stellar activity signatures in the light curves of FGK stars are often weak compared to M dwarfs \citep{SurezMascareo2017}, making clear rotational modulation uncommon in our sample. We do not re-fit the transit light curves to improve planetary radii precision, as the primary source of heterogeneity in planetary bulk density measurements often arises from the RV determination of planetary masses \citep{Osborne2024}. Planetary radii derived from Kepler, K2, and TESS are based on homogeneous, well-characterised space-based observations processed through dedicated survey pipelines \citep{Jenkins2010, Jenkins2016}, whereas RV analyses often differ substantially in data selection, reduction, activity treatment, and statistical methodology. A fully homogeneous re-analysis of the transit photometry and stellar parameters is beyond the scope of this work.

\subsection{Stellar rotation}\label{stellar_rotation}

We searched for stellar rotation signals in both space-based photometry and spectroscopic activity indicators derived from the RV datasets, restricting all searches to $P \in [2,100]$\,d, a lower bound that admits fast-rotating young FGK stars while staying clear of the 1\,d sampling alias. We analysed each available time series with the Generalised Lomb--Scargle periodogram \citep[GLS,][]{ZechmeisterKurster2009}, whose peak significance is quantified by the false alarm probability (FAP), computed from the single-frequency probability,
\begin{equation}
\mathrm{FAP}(z) = (1-z)^{(N-3)/2}
\label{eq:fap_single}
\end{equation}
for a periodogram power $z$ from $N$ points, generalised to the full frequency range via the analytic alias correction of \citet{Baluev2008}.

We applied the GLS independently in the photometric and spectroscopic domains, yielding two period measurements per target. For photometry, the periodogram acts directly on the light curve. For spectroscopy, we drew on activity indicators from each available reduction: the differential line-width (dLW) from LBL \citep{Artigau2022}, the dLW from SERVAL and sBART \citep{Zechmeister2017,Silva2022}, the CCF full width at half maximum (FWHM), contrast, and Na and Ca line indices \citep{Baranne1996}, and the chromospheric $S$-index from the HIRES/NEID/APF data products \citep{Duncan1991}. A per-indicator detection was retained only if $N\geq12$ (where $N$ is the number of RV or photometric measurements for that indicator), $P\in[2,100]\,\mathrm{d}$, and $P<T_\mathrm{base}/2$ (where $T_\mathrm{base}$ is the total time baseline of the dataset). The retained detections were combined into a single measurement per domain, with each indicator-level detection assigned a weight via
\begin{equation}
    w = \sqrt{1 + \mathrm{S/N}} \times s,
    \label{eq:detection_weight}
\end{equation}
where
\begin{equation}
\mathrm{S/N} = \tfrac{1}{2}(P_{95}-P_{5})/\sigma_\mathrm{noise}.
\label{eq:snr}
\end{equation}
Here, $P_5$ and $P_{95}$ are the 5th and 95th percentiles of the indicator values, $\sigma_\mathrm{noise}$ is the median error, and $s = -\!\log_{10}\mathrm{FAP}$ is the significance score. The S/N is capped at 50, so that an indicator with an underestimated noise floor cannot end up dominant. The domain rotation period $P$ is the weighted median of these detections, with $\sigma_P$ the largest of the median per-indicator formal error, $1.4826$ times the weighted median absolute deviation of the periods about $P$, and a 20\,\% floor following \citet{Figueira2025}. The $P_\mathrm{rot}$ prior was taken from the photometric estimate wherever one reached $\mathrm{FAP}<10^{-3}$ and from the spectroscopic indicators otherwise, photometry being the more direct tracer of rotational modulation (Table~\ref{tab:prot_summary}), which every target with a detection satisfied.

\begin{table}[htbp]
\centering
\footnotesize
\setlength{\tabcolsep}{3pt}
\caption{Derived stellar rotation periods used as prior for RV modelling.}
\label{tab:prot_summary}
\begin{tabular}{lcccc}
\hline
Target & $P_\mathrm{rot}$ (d) & $P_\mathrm{rot}^\mathrm{lit}$\tablefootmark{a} (d) & Ref. & Amp.\tablefootmark{b} \\
\hline
K2-182     & $26.00 \pm 5.20$  & $24.84 \pm 3.69$ & Rei20 & 4.59 \\
K2-184     & $20.91 \pm 4.18$  & $16.67 \pm 1.35$ & Rei20 & 1.40 \\
K2-244     & \ldots            & \ldots           & \ldots & 0.09 \\
K2-269     & $11.84 \pm 2.37$  & $12.40 \pm 0.80$ & Rei20 & 0.22 \\
K2-66      & $27.24 \pm 10.33$ & \ldots           & \ldots & 0.54 \\
Kepler-126 & $7.21 \pm 1.44$   & $8.11$           & Rei23 & 0.12 \\
Kepler-129 & $10.02 \pm 2.00$  & $29.78 \pm 0.56$ & Maz15 & 0.15 \\
Kepler-145 & $13.53 \pm 2.71$  & $13.53$          & Rei23 & 0.11 \\
Kepler-21  & $12.65 \pm 2.53$  & $12.62 \pm 0.03$ & Lop16 & 0.21 \\
Kepler-538 & $46.92 \pm 9.38$  & $24.10 \pm 4.58$ & San21 & 0.61 \\
Kepler-65  & $8.03 \pm 1.61$   & $7.91 \pm 0.15$  & McQ13 & 0.14 \\
TOI-1411   & $8.01 \pm 1.60$   & \ldots           & \ldots & 0.47 \\
\hline
\end{tabular}
\tablefoot{
Reference abbreviations:
Rei20 = \citet{Reinhold2020},
Rei23 = \citet{Reinhold2023},
Maz15 = \citet{Mazeh2015},
Lop16 = \citet{LopezMorales2016},
San21 = \citet{Santos2021},
McQ13 = \citet{McQuillan2013}.
K2-244 yields no GLS detection reaching $\mathrm{FAP}<10^{-3}$ in either domain and is given an uninformative $\eta_3$ prior.
\tablefootmark{a}One published value per star, preferring direct photometric measurements. The compilation is not exhaustive. Values are quoted without uncertainty where the source gives none.
\tablefootmark{b}Half the 5th--95th percentile range of the de-trended, de-transited light curve in ppt, the amplitude term of Eq.~\ref{eq:snr}.}
\end{table}

Our derived rotation periods agree with the published values for six of the nine targets that have them. For Kepler-129 our period is one third of the published value and for Kepler-538 about twice it, so both are most likely harmonics of the published period, and for K2-184 the two published values come from separate K2 campaigns of the same star, disagree with each other by 30\,\%, and bracket ours without confirming it. We report these disagreements rather than resolve them. None of the three adopts a GP in its final noise model (Sect.~\ref{sec:modelling}), so none contributes a rotation prior to any mass reported here.

\subsection{Magnetic cycles}

We searched for long-baseline ground-based photometric time series for all targets in our sample in the All-Sky Automated Survey for Supernovae \citep[ASAS-SN;][]{Kochanek2017} and the Zwicky Transient Facility \citep[ZTF;][]{Bellm2019}.
The Catalina Sky Survey (CSS), the Super Wide Angle Search for Planets (SuperWASP), and the American Association of Variable Star Observers (AAVSO) were queried, but no matches were found. We also checked the Hungarian-made Automated Telescope Photometric Survey (HATPI; \citealt{Hartman2026}), whose Las Campanas footprint covers only the five K2 targets in our sample. None of K2-182, K2-184, K2-269, and K2-66 shows a significant periodicity, while K2-244's 179\,d baseline is too short for a long-period search. ASAS-SN photometry was retrieved via the Sky Patrol v2 interface \citep{Hart2023}. The survey comprises two distinct photometric phases: $V$-band (2013--2018; Phase~1) and $g$-band (2017--present; Phase~2). In our sample, usable data were recovered only from Phase~1. ZTF light curves were retrieved from the NASA/IPAC Infrared Science Archive (IRSA) light-curve service using a $2\arcsec$ cone search, with flagged epochs (\texttt{catflag} $= 32768$) removed.

We retained datasets with at least 20 points and epochs with photometric uncertainties $\sigma_{\rm mag} < 0.1$\,mag. For ASAS-SN, the shortest light-curve baseline in our sample is approximately 1700\,d, with a median cadence of $\sim$3\,d and reported photometric uncertainties of $\sim$20\,mmag, which makes ASAS-SN the primary source by target coverage and retained points. Of the 12 targets in the sample, 11 have usable ASAS-SN photometry and 3 have usable ZTF photometry, while TOI-1411 returned no data in any of the queried databases. Period searches were performed independently for each survey using the Lomb--Scargle periodogram.  The frequency grid was set automatically according to the data sampling.

We searched for periodicities between $P_{\rm min} = 20$\,d and $P_{\rm max} = 0.8 \times T_{\rm base}$, where $T_{\rm base}$ is the total baseline of each dataset, so that at least $\sim$1.25 cycles fit in the observation window. The statistical significance of each peak was assessed using the FAP. No significant periodicity (FAP $< 0.05$) attributable to the stellar rotation period was recovered in any of the targets. This null result is consistent with the low photometric amplitudes expected for mature solar-type stars, but for the shorter and more sparsely sampled light curves in the sample it may equally reflect limited sensitivity rather than genuine quiescence. Three targets show marginal periodicities at the 1--3\% FAP level, consistent with seasonal aliasing (most prominently at $P \sim 365$\,d and its harmonics). The seasonal mean magnitudes of K2-184, Kepler-21, Kepler-65, and Kepler-126 show a statistically significant ($>5\sigma$) linear trend, with amplitudes $\Delta m = 0.06$--$0.10$\,mag over the available baselines. We cannot tell whether these trends reflect long-term magnetic cycles or low-level instrumental systematics. Only Kepler-21 adopts a linear RV trend (Sect.~\ref{sec:noise_model_selection}), and for this star the long-term RV variability may be stellar or due to the outer companion candidate Kepler-21\,c \citep{BeardRobertson2024}.

\section{RV modelling and inference} \label{sec:modelling}

The RV analysis was performed using the Bayesian inference package \texttt{kima}\footnote{\url{https://www.github.com/kima-org/kima}} \citep{Faria2018} \texttt{v6.3.0}, which implements a trans-dimensional framework specifically designed for exoplanet detection in RV data \citep{Brewer2015}. In contrast to traditional approaches that fix the number of planets a priori, \texttt{kima} treats the number of Keplerian signals, $N_{\rm p}$, as a free discrete parameter inferred directly from the data. This allows the sampler to explore models of different dimensionality within a single inference run, providing posterior probabilities for competing planetary architectures. Parameter exploration and evidence computation were performed using diffusive nested sampling (DNS; \citealt{Brewer2010}), which simultaneously delivers posterior distributions and the Bayesian evidence ($\ln Z$). It then estimates the most probable planetary configuration from the relative fraction of posterior samples between $N_p$ bins.

For each star, the observed RVs are modelled as the sum of $N_{\rm p}$ Keplerian signals together with instrumental offsets, additional white-noise terms, and a polynomial trend (up to an order of two) accounting for long-term variability. The polynomial component absorbs the reflex acceleration of unresolved long-period companions, as well as possible stellar magnetic-cycle variability. The polynomial order was selected through Bayesian model comparison using the marginal likelihood returned by \texttt{kima}. This strategy allows long-term signals to be incorporated only when supported by the data while avoiding overfitting. Each instrument was assigned an independent RV offset and jitter term. We fitted two sets of models, one containing only the known planets, from which all masses were measured, and one with additional free Keplerians, used to search for new companions (Sect.~\ref{sec:free_planet_search}). Priors are given in Sect.~\ref{sec:ko_priors}.

\subsection{GP modelling of stellar activity and model selection}

Stellar activity signals are modelled using the quasi-periodic (QP) Gaussian-process kernel implemented natively in \texttt{kima}. This kernel has become a standard choice for modelling rotationally modulated activity in RV datasets \citep{Stock2023} and captures the quasi-periodic evolution of active regions on stellar surfaces. The QP kernel takes the form
\begin{equation}
    k(t_i, t_j) = \eta_1^2 \exp\!\left[-\frac{(t_i - t_j)^2}{2\eta_2^2} - \frac{2\sin^2\!\left(\pi(t_i - t_j)/\eta_3\right)}{\eta_4^2}\right],
\end{equation}
where $\eta_1$ is the GP amplitude, $\eta_2$ is the evolutionary (decay) timescale of active regions, $\eta_3$ is the stellar rotation period, and $\eta_4$ is the smoothing factor governing the harmonic complexity of the activity signal \citep{Haywood2014}. Because stellar activity levels vary significantly across the targets, GP modelling is not imposed universally. Instead, its inclusion was determined through Bayesian model comparison. The GP hyperparameter posteriors for the targets whose adopted model includes a GP are reported in Table~\ref{tab:gp_hyper}.

For each system, we performed a systematic exploration of model complexity through a two-dimensional grid of models combining polynomial trends of degree 0, 1, and 2, and models with or without a Gaussian-process activity component. This results in six competing noise models for each star. Models are ranked according to increasing complexity, where the inclusion of a GP represents the largest increase in flexibility, followed by higher-order polynomial trends. A more complex model was adopted only when its evidence exceeded that of simpler alternatives by $\Delta \ln Z \geq 5$, corresponding to a Bayes factor of $\mathrm{BF} \approx \mathrm{e}^5 \approx 148$. For multi-reduction targets, $\Delta \ln Z$ is the mean over reductions of the evidence difference within each reduction, and the same threshold applies to every target. This hierarchical approach ensures that additional model components are introduced only when statistically justified, providing a uniform treatment of stellar variability across the sample. The full grid of evidence differences for all six models and all 12 targets is given in Table~\ref{tab:dlnz}.

\subsection{Leave-one-out cross-validation and Bayesian model averaging across RV reductions}
\label{sec:loocv_bma}

In essence, leave-one-out cross-validation (LOO-CV) asks how well each fitted model predicts each of its own data points when that point is withheld, and rewards models whose predictions hold up best. This gives a principled, data-driven way to weight the four independent RV extractions available for each target, rather than selecting a single reduction a priori.

Different RV extraction pipelines (CCF, SERVAL, LBL, sBART) encode slightly different assumptions about spectral information extraction and noise, and therefore deliver slightly different posterior distributions for the same physical parameters. To combine posteriors obtained from different reductions, we assigned weights $\{w_r\}$ from LOO-CV, which scores how well each reduction's dataset predicts its own observations under a fixed astrophysical model. For four targets observed exclusively with HIRES or NEID (K2-66, Kepler-129, Kepler-145, and TOI-1411), no multi-pipeline extraction was available and the LOO-CV stacking framework was not applied. The reported masses for these targets were derived directly from the available RVs. 

For a given target and fixed astrophysical model $M$, let $r$ index the reduction method and let $\mathbf{y}_r$ denote the RV time series from that reduction (with aligned epochs and a common observation count across reductions). The expected log pointwise predictive density under LOO is
\begin{equation}
\mathrm{elpd}_{\mathrm{LOO},r}
  = \sum_{i} \log p\bigl(y_{r,i} \mid \mathbf{y}_{r,-i}, M\bigr),
\label{eq:elpd_loo}
\end{equation}
where $y_{r,i}$ is the $i$-th RV measurement in reduction $r$ and $\mathbf{y}_{r,-i}$ denotes all remaining points. A larger $\mathrm{elpd}_{\mathrm{LOO},r}$ indicates that the observations in reduction $r$ are more mutually predictable under $M$, meaning each point is well-described by the model fit to the rest. Evaluating \eqref{eq:elpd_loo} requires, for each $i$, the marginalised LOO predictive over parameter space,
\begin{equation}
   p(y_{r,i} \mid \mathbf{y}_{r,-i}, M) = \int p\bigl(y_{r,i} \mid \mathbf{y}_{r,-i}, \boldsymbol{\theta}, M\bigr)\, p(\boldsymbol{\theta} \mid \mathbf{y}_{r,-i}, M)\,\mathrm{d}\boldsymbol{\theta},
\end{equation}
which is expensive if computed by explicit refits. We proceeded in two conceptual steps. First, for each posterior draw $\boldsymbol{\theta}_s \sim p(\boldsymbol{\theta} \mid \mathbf{y}_r, M)$ we formed the pointwise log predictive score $\ell_{r,s,i} = \log p\bigl(y_{r,i} \mid \mathbf{y}_{r,-i}, \boldsymbol{\theta}_s, M\bigr)$. When observations are conditionally independent given $\boldsymbol{\theta}$, which is the case for white-noise models, $\ell_{r,s,i}$ is just the ordinary per-point log-likelihood. When $M$ includes a GP residual model, correlations couple the points and we cannot consider them independent. We then used the standard closed-form LOO identities for multivariate Gaussian observations \citep[Sect.~5.4.2, eqs.~5.10--5.12]{Rasmussen2006} so that $\ell_{r,s,i}$ is still obtained without refitting the GP separately for each $i$. Second, we approximated the integral over $\boldsymbol{\theta}$ by Pareto-smoothed importance sampling LOO (PSIS-LOO), which reweights the posterior draws $\boldsymbol{\theta}_s$ towards the leave-one-out posterior for each $i$ and stabilises the importance weights by smoothing their upper tail with a fitted generalised Pareto distribution. This yields a pointwise contribution of $\widehat{\mathrm{elpd}}_{\mathrm{LOO},r,i}$ that sums to a scalar $\widehat{\mathrm{elpd}}_{\mathrm{LOO},r}$ for each reduction. Diagnostic Pareto shape parameters $k_i$ flag observations for which the PSIS approximation is unreliable. We chose to reject the approximation when $k_i>0.7$ \citep{Vehtari2016}. For these flagged observations, we needed to re-run the model without them to obtain their true $\mathrm{elpd}_{\mathrm{LOO},r,i}$ values.

We converted the predictive performance into importance weights for each reduction $\{w_r\}$ with $\sum_r w_r = 1$ in two ways. We first computed the pseudo-Bayesian model averaging (pseudo-BMA) metric, which sets weights that are proportional to $\exp\bigl(\widehat{\mathrm{elpd}}_{\mathrm{LOO},r}\bigr)$ equivalent to normalising softmax weights over reductions. Then, we computed the stacking metric, which selects $\{w_r\}$ to maximise $\sum_i \log \sum_r w_r \, \exp\bigl(\widehat{\mathrm{elpd}}_{\mathrm{LOO},r,i}\bigr)$ \citep{Yao2018}. Since all reductions derive from the same spectra, the datasets are not independent. Stacking was preferred here as it allowed us to optimise the predictive score of a mixture of reductions rather than treating reductions as mutually independent alternatives. In addition, stacking mitigates over-weighting effects when several reductions are very similar to one another. For instance, SERVAL and sBART are conceptually close, as both rely on masking strategies for telluric contamination and order-by-order template matching for RV extraction. Treating such closely related reductions as independent alternatives would artificially inflate their collective influence. This general approach has also been applied in exoplanet atmospheric analysis to combine multiple reductions or retrievals across different models \citep{Welbanks2023,Nixon2024}.

For parameters $\boldsymbol{\phi}$ shared across reductions, the LOO-weighted Bayesian model average is again a mixture of reduction-specific posteriors,
\begin{equation}
p(\boldsymbol{\phi} \mid \{\mathbf{y}_r\}, M)
  = \sum_r w_r \, p(\boldsymbol{\phi} \mid \mathbf{y}_r, M),
\label{eq:loocv_bma_mixture}
\end{equation}
where $p(\boldsymbol{\phi} \mid \mathbf{y}_r, M)$ is the marginal posterior from reduction $r$. The final posteriors used in this study were obtained from the stacking mixture of the four reductions mentioned in Sect.~\ref{rv_extraction_methods}. The resulting pipeline weights are discussed in Sect.~\ref{sec:pipeline_comparison}.

Despite the terminology, this is not a fully Bayesian model average: a strict version would require each reduction's marginal likelihood and a prior over the reductions; neither is well defined when the four reductions are correlated realisations of the same spectra, rather than independent competing models. Instead, the pseudo-BMA and stacking weights optimise the predictive accuracy directly via LOO-CV.

\subsection{Adopted priors}
\label{sec:ko_priors}

Systemic velocity, offsets, jitter, and trend coefficients use \texttt{kima}'s default priors, uninformative on the scale of the data (offsets and jitter scale with $K_\mathrm{max}$, the peak-to-peak span of the RV data). The GP hyperparameters take \texttt{kima}'s default log-uniform priors: $\eta_1 \sim \mathcal{U}_\mathrm{log}(10^{-5}\sigma_1,\,10\,\sigma_1)\,\mathrm{m\,s^{-1}}$, with $\sigma_1$ the RV scatter of the reference instrument, $\eta_2 \sim \mathcal{U}_\mathrm{log}(5,\,500)\,\mathrm{d}$, and $\eta_4 \sim \mathcal{U}_\mathrm{log}(0.1,\,10)$. The evolution timescale is constrained to $\eta_2 > \eta_3$, so that active regions cannot decay faster than the star rotates, and the prior on the rotation timescale $\eta_3$ is informed by the analysis of Sect.~\ref{stellar_rotation} (Table~\ref{tab:prot_summary}).

Each known planet has a log-uniform prior on its semi-amplitude, $K \sim \mathcal{U}_\mathrm{log}(0.1,\,K_\mathrm{max})\,\mathrm{m\,s^{-1}}$, a Kumaraswamy$(0.867,\,3.03)$ prior on its eccentricity, which reproduces the empirical eccentricity distribution of the known exoplanet population \citep{Kipping2013}, and a uniform prior on $\omega \in [0,\,2\pi)$. The transiting planets have Gaussian priors on their orbital period and time of transit from the literature ephemeris. The non-transiting giants Kepler-65\,e and Kepler-129\,d were discovered in RVs that we fitted as well, so they have uniform priors on orbital phase and, over $244.8$--$272.8$\,d for Kepler-65\,e and $1500$--$3267$\,d for Kepler-129\,d, on period, to avoid double-dipping. The first window is centred on the published $258.8$\,d with a half-width of 14\,d. The second contains both published periods \citep{Zhang2021,Weiss2023} and ends at the HIRES baseline. Free planets, used for the search of Sect.~\ref{sec:free_planet_search}, have a uniform prior on period, a log-uniform prior on semi-amplitude, and circular orbits.

\subsection{Search for additional companions}
\label{sec:free_planet_search}

The masses we report come from fits that contain only the known planets, but to search for additional companions, we exploited the trans-dimensional nature of \texttt{kima} and allowed up to $N_\mathrm{p}^\mathrm{max} = 2$ free planets per target. Additional-planet orbits were held circular since eccentric models introduce sufficient additional flexibility that even a Jeffreys prior on $e$ does not reliably prevent the sampler from fitting noise features and inflating the evidence for spurious signals \citep{Lucy1971,Zakamska2010,Hara2019}. The posterior samples were labelled by $N_\mathrm{p}$, and the logarithmic Bayes factor comparing adjacent architectures follows directly from the sample counts,
\begin{equation}
  \ln B_{N,\,N-1} = \ln \frac{n_N}{n_{N-1}},
  \label{eq:bayes_factor}
\end{equation}
where $n_N$ and $n_{N-1}$ are the nested-sampling posterior weights in the $N_\mathrm{p} = N$ and $N_\mathrm{p} = N-1$ bins, respectively. If the $N_\mathrm{p} = N-1$ bin is empty, its probability is below $3/n_\mathrm{tot}$ at 95\% confidence for $n_\mathrm{tot}$ posterior samples, and we used the resulting lower bound $\ln B_{N,\,N-1} \geq \ln(n_N/3)$.
A free signal is a detection only if it passes four vetting criteria in consecutive order: a decisive Bayes factor ($\ln B_{1,0}>5$, Tier~1); well-behaved posterior morphology ($K/\sigma_K > 3$; amplitude skewness $\gamma_1(K) < 1$, and at least 90\,\% of period samples within $\pm 5\,\%$ of the maximum a posteriori (MAP) period, Tier~2); a decisive detection with consistent periods (within $5\,\%$) in at least two reductions (Tier~3, which only applies to targets with several reductions); and no match within $5\,\%$ to the solar year or its first harmonic (Tier~4).

\section{Results and discussion} \label{sec:results}

\subsection{Reduction comparison and LOO stacking}
\label{sec:pipeline_comparison}

\begin{figure*}[!h]
\centering
\includegraphics[width=\textwidth,height=0.7\textheight,keepaspectratio]{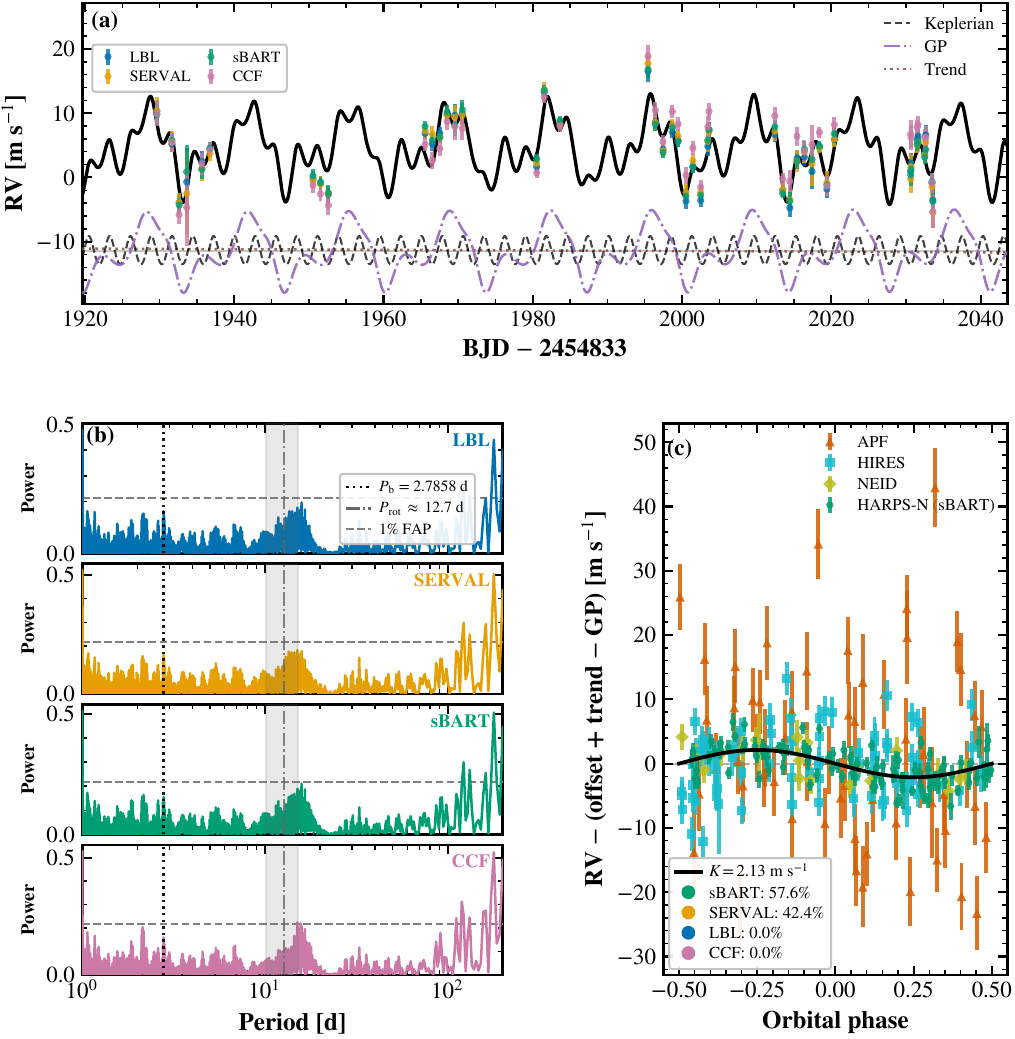}
\caption{RV analysis of Kepler-21\,b. (a) HARPS-N RVs over the densest 130\,d window for each pipeline, with the model (Keplerian + GP + linear trend) evaluated on the best-weighted reduction (sBART). (b) GLS periodograms per pipeline. (c) RVs phase-folded on $P_b$ after subtracting offsets, trend, and GP. The planet is detected at $7.1\sigma$ ($K = 2.13$\,m\,s$^{-1}$) alongside a quasi-periodic activity signal at $P_\mathrm{rot} \approx 12.7$\,d.}
\label{fig:kepler21_rv_hero}
\end{figure*}

For each multi-pipeline target, we fitted the RVs of every reduction separately and combined the resulting posteriors as a weighted mixture. The weights were chosen by stacking, so that the mixture can best predict each RV when that RV is left out of the fit, a score we estimated with PSIS-LOO (Sect.~\ref{sec:loocv_bma}). Thus, a reduction whose model predicts held-out RVs poorly contributes little to the final mass. No single reduction pipeline is optimal across the sample, while the LOO-CV stacking weights make the division of labour explicit. Figure~\ref{fig:kepler21_rv_hero} shows the analysis of Kepler-21 as a worked example, and Appendix~\ref{app:pipelines} the per-pipeline fits. 

\begin{figure*}
\centering
\includegraphics[width=\textwidth]{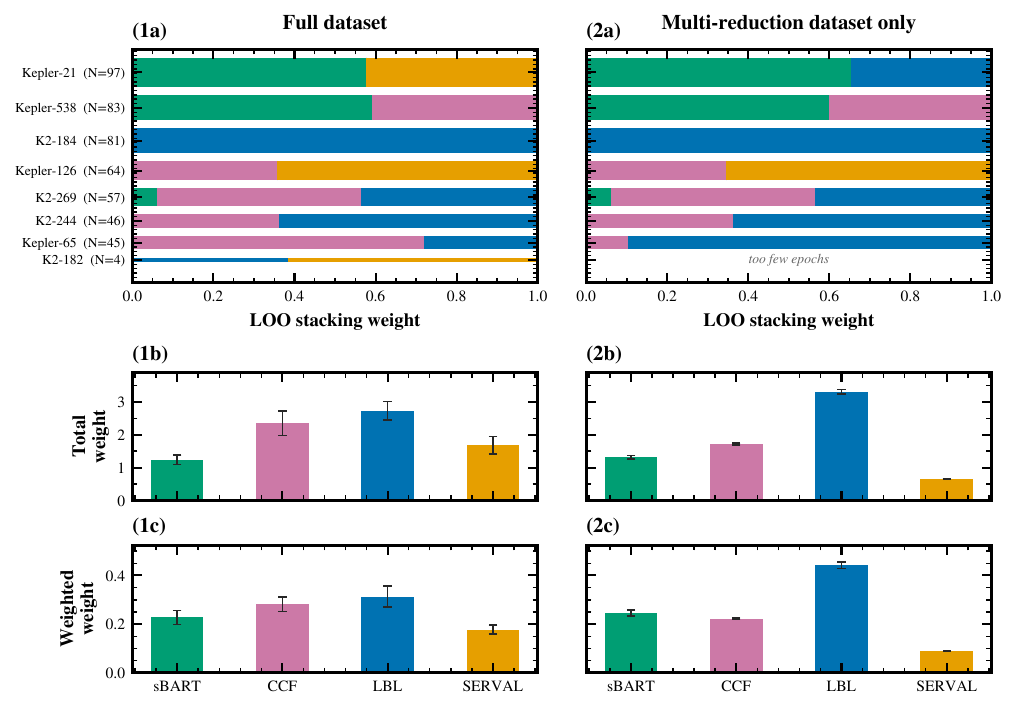}
\caption{LOO-CV stacking weights per pipeline and target. Column 1: full dataset, whose weights give the final posteriors. Column 2: multi-reduction dataset (HARPS-N and ESPRESSO only). Row a: per-target weights, with bar thickness scaling with the number $N$ of HARPS-N and ESPRESSO RVs (in parentheses). Row b: total weight per pipeline, summed over the eight (column 1) and seven (column 2) targets. Row c: mean weight per pipeline, weighting each target by $N$. Error bars are the standard deviation over five sampler seeds. Only K2-184 is dominated by a single pipeline (\texttt{LBL}), and no pipeline is optimal across the sample.}
\label{fig:loo_weights}
\end{figure*} 

The per-target weights (Fig.~\ref{fig:loo_weights}, row a) confirm that the stacking genuinely mixes information for most systems. Only K2-184 is dominated by a single pipeline, with all its weight on \texttt{LBL}, and the remaining seven multi-pipeline targets spread their weight across two or three reductions. Stacking barely shifts the central mass estimates. For the 14 planets of the eight multi-reduction targets, the median semi-amplitudes of the four single pipelines span a median of $0.75\,\sigma_K$ of the stacked posterior (Fig.~\ref{fig:mosaic_reduction}). The span is below $1.3\,\sigma_K$ for 13 of them and reaches $2.7\,\sigma_K$ for Kepler-126\,c. At least for this sample, the multi-pipeline approach buys robustness against an unlucky single-reduction choice, not a systematic shift in the derived masses. We also tested whether the weight each pipeline receives depends on $T_\mathrm{eff}$, [Fe/H], or $P_\mathrm{rot}$ with linear fits that account for the uncertainties on both axes and on the weights, and found that a sample of eight targets is too small to yield any slope inconsistent with zero at 95\% credibility. A larger sample should test these dependencies.

These weights were fitted on all the RVs of a target, including the HIRES, APF, and NEID velocities, which enter every per-reduction fit unchanged. For Kepler-21, Kepler-65, Kepler-126, and Kepler-538, the weights therefore partly measure how each reduction agrees with the RVs of the single-reduction instruments. To test the stacking without this external contribution, we repeated the analysis on a multi-reduction dataset, which keeps only the HARPS-N and ESPRESSO RVs reduced by all four pipelines, for every multi-pipeline target except K2-182. Its four HARPS-N RVs span only two nights. On the multi-reduction dataset (Fig.~\ref{fig:loo_weights}, column 2) the stacked semi-amplitude agrees with the single-pipeline results for all seven targets and in every seed. Its 95\% interval contains the median of each pipeline that is not rejected, and its median lies within $1\sigma$ of theirs. The weights are also reproducible, with a seed-to-seed standard deviation below 0.08 for every pipeline and target. On the full dataset only Kepler-21 and Kepler-65 have weights that scatter by more than 0.1 between seeds. For Kepler-65, the HIRES RVs pin the period of the 259-d giant, and its eccentricity posterior then has two modes, one at $e\approx0.25$ and one at low $e$ with unconstrained $\omega$. The nested sampler of the \texttt{CCF} fit finds the first in only one of five seeds, which moves the stacking weights and the stacked $K_e$ by up to $0.37\,\sigma_K$.

Adding the HIRES, APF, and NEID RVs changes the weights of two of the four targets that have them, Kepler-21 and Kepler-126, by more than their seed scatter. For Kepler-21, in the final posteriors, the weight moves from \texttt{sBART} (65\%) and \texttt{LBL} (35\%) on the multi-reduction-only run to \texttt{sBART} (58\%) and \texttt{SERVAL} (42\%) on the full dataset. The semi-amplitudes move much less. The shift between the two datasets is below the $1\sigma$ width of the full-dataset posterior for eight of the ten planets and reaches $1.6\sigma$ for Kepler-126\,b and $1.3\sigma$ for Kepler-538\,b. The stacked masses therefore depend little on the inclusion of the single-reduction instruments, while for some targets the weights depend on it strongly. Since the full-dataset weights also score the agreement with the HIRES, APF, and NEID RVs, whose own accuracy we cannot verify, we do not read them as a ranking of pipeline quality. The multi-reduction weights (second column of Fig.~\ref{fig:loo_weights}) come closer to such a ranking, because each reduction is scored only on held-out RVs it produced itself. Seven targets, most of whose planets yield only upper limits, are still too few to rank the pipelines, and that comparison should be made on a larger sample with a high S/N. 

\subsection{Noise model selection across the sample}
\label{sec:noise_model_selection}
Of the six noise models compared for each target (Sect.~\ref{sec:modelling}), two targets required a GP. For Kepler-21 the GP is preferred by $\Delta\ln\mathcal{Z}=98.3$ over its best model without one. For K2-66 a linear trend is first preferred over a constant offset ($\Delta\ln\mathcal{Z}=38.0$) and the GP then improves on it by 6.7. The remaining ten targets show sub-threshold support for a GP, Kepler-538 and K2-184 coming closest at $\Delta\ln\mathcal{Z}=3.9$ and 2.6 (Table~\ref{tab:dlnz}). Every adopted model has a constant instrumental offset except for Kepler-21, which supports a linear trend alongside the GP. This is consistent with both rotation-modulated activity and a slow RV drift, which \citet{BeardRobertson2024} identify as the long-period candidate Kepler-21\,c. The GP hyperparameters of Kepler-21 and K2-66 are reported in Table~\ref{tab:gp_hyper}.

\subsection{Planetary masses and their compositional implications}
\label{sec:masses}

\begin{figure}[!t]
\centering
\includegraphics[width=\columnwidth]{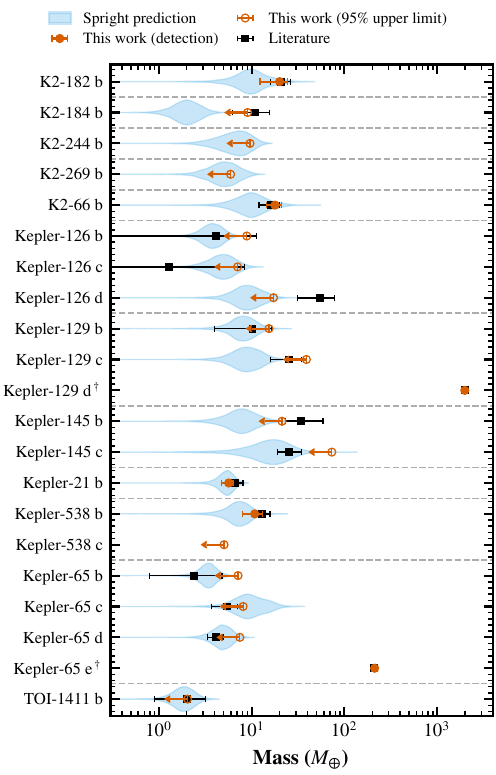}
\caption{Planetary masses, $M_p$, for the 21 planetary signals in the sample ($M_p\sin i$ for the non-transiting giants, marked $\dagger$), compared with literature masses and \textsc{spright} predictions from the planet radius (none for Kepler-538\,c, whose radius lies below the range of the relation). Six signals are detections and fifteen are 95\% upper limits. The four transiting detections lie at or above the radius-based predictions, and the literature masses are recovered to within $2.1\sigma$.} 
\label{fig:mass_constraints}
\end{figure}

Table~\ref{tab:masses} lists the full planetary-parameter set, and Fig.~\ref{fig:mass_constraints} summarises the mass constraints for the full sample. Of the 21 planetary signals, six exceed the $3\sigma$ detection threshold. Four are transiting planets, Kepler-21\,b ($K/\sigma_K = 7.1$), K2-66\,b (5.5), Kepler-538\,b (4.2), and K2-182\,b (3.3), while two are the non-transiting giants Kepler-129\,d (61.1) and Kepler-65\,e (14.7). The remaining fifteen signals yield upper bounds on planetary mass. Two systems (K2-244 and K2-269) had no prior mass measurement and this work provides their first RV constraints, though neither reaches a $3\sigma$ detection.

Figure~\ref{fig:mass_constraints} compares the mass constraints with their radius-based \textsc{spright} priors and the literature masses $M_p^{\rm lit}$ of Table~\ref{tab:masses}. For the systems with prior mass determinations the stacked mass posteriors agree with the literature to within $1$--$2\sigma$. Kepler-21\,b gives $5.66^{+0.65}_{-0.96}\,\mearth$, against the RV mass of $6.68 \pm 1.4\,\mearth$ \citep{BeardRobertson2024} and the earlier HARPS-N RV mass of $7.5 \pm 1.3\,\mearth$ \citep{Bonomo2023}. K2-66\,b gives $17.9 \pm 3.2\,\mearth$, against the HIRES RV mass of $16 \pm 4\,\mearth$ \citep{Howard2025}. K2-182\,b gives $20.1^{+4.5}_{-7.8}\,\mearth$, matching the HIRES RV mass of $21.0^{+5.4}_{-5.1}\,\mearth$ \citep{Thygesen2023}, and Kepler-538\,b gives $10.8^{+2.4}_{-2.8}\,\mearth$ against the HARPS-N RV mass of $12.9 \pm 2.9\,\mearth$ \citep{Bonomo2023}. The detection of K2-182\,b is seed-sensitive ($K/\sigma_K$ between 2.6 and 3.5 over five sampler seeds), and that of Kepler-538\,b depends on the noise model, since a GP lowers its semi-amplitude to $0.2$--$0.4\,\mathrm{m\,s^{-1}}$. K2-184\,b gives $<\!9.1\,\mearth$ against the ESPRESSO mass of $10.9 \pm 4.7\,\mearth$ \citep{Hobson2026}, whose semi-amplitude, $4.4^{+1.5}_{-1.6}\,\mathrm{m\,s^{-1}}$, lies $1.9\sigma$ above ours ($K<2.7\,\mathrm{m\,s^{-1}}$ at 95\%). The two analyses largely share the same ESPRESSO spectra, so they are not independent. \citet{Hobson2026} fit an eccentric orbit ($e=0.57$) and a GP jointly with the K2 photometry, whereas our adopted model has no GP and an eccentricity set mostly by its prior, and we add the HARPS-N spectra.

\begin{figure*}
    \centering
    \includegraphics[width=\textwidth]{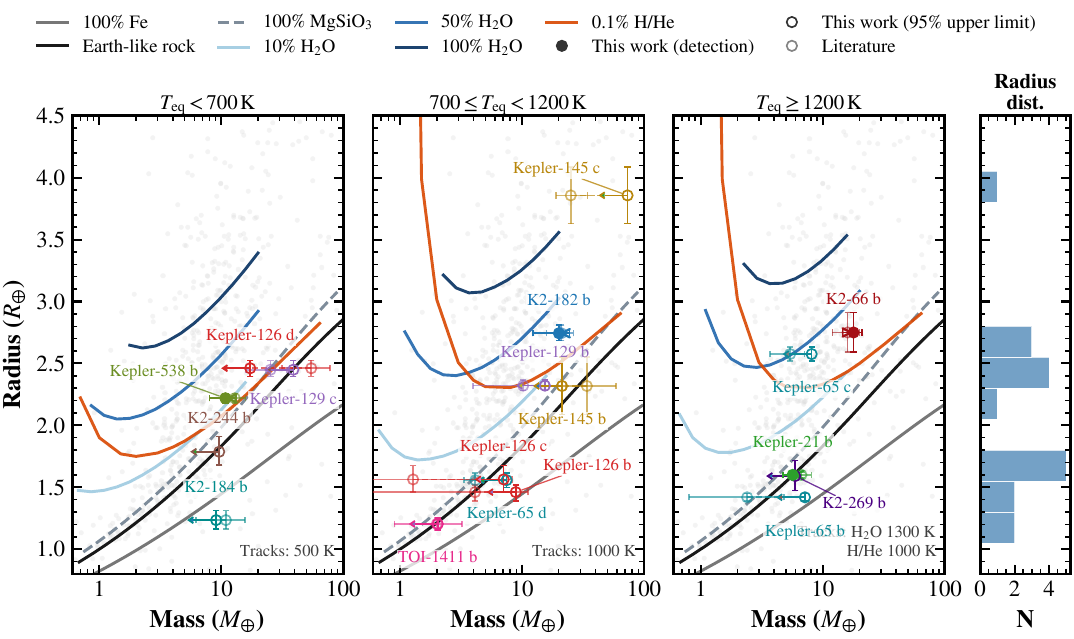}
    \caption{Mass--radius diagram in three equilibrium-temperature bins (cool: $T_{\rm eq}<700$\,K; warm: $700$--$1200$\,K; hot: $\geq1200$\,K) and radius distribution of the sample (right). Detections (filled) and 95\% upper limits (open, leftward arrows) are joined to the literature masses (faint circles). Kepler-21\,b lies on the Earth-like rock track, whereas K2-182\,b, K2-66\,b, and Kepler-538\,b are too large for a rocky interior. Planet radii are the published $R_p/R_\star$ multiplied by the adopted $R_\star$ of Table~\ref{tab:sample}, except for K2-66\,b, whose radius is taken from \citet{Howard2025}. Grey background: NASA Exoplanet Archive (accessed 2026-07-21). Rocky tracks are from \citet{Zeng2019}, and the irradiated ocean-world and $0.1\%$\,H/He envelope tracks from \citet{Aguichine2021}, at the temperatures noted in each panel.}
    \label{fig:mr_diagram}
\end{figure*}

Figure~\ref{fig:mr_diagram} places the transiting planets on the mass--radius diagram. The four transiting detections span bulk densities from $4.7\,\mathrm{g\,cm^{-3}}$ for K2-66\,b to $7.6\,\mathrm{g\,cm^{-3}}$ for Kepler-21\,b. The radius of Kepler-21\,b is consistent with a rocky interior of Earth-like composition and no envelope. The other three are too large for a rocky interior, lying above the 100\% MgSiO$_3$ track, the least dense rocky model, at their median masses. They require a volatile component: a water mass fraction between 10\% and 50\% for K2-66\,b ($4.69\,\mathrm{g\,cm^{-3}}$) and K2-182\,b ($5.33\,\mathrm{g\,cm^{-3}}$) and close to 10\% for Kepler-538\,b ($5.47\,\mathrm{g\,cm^{-3}}$), or an H/He envelope above 0.1\% of the planet mass. Mass and radius alone cannot distinguish these two families.

The two non-transiting giants, Kepler-129\,d ($M_p\sin i = 1986^{+29}_{-35}\,\mearth$, against the HIRES $M_p\sin i$ of $2015 \pm 43\,\mearth$; \citealt{Weiss2023}) and Kepler-65\,e ($M_p\sin i = 212^{+11}_{-14}\,\mearth$, against $207.5^{+17.8}_{-17.5}\,\mearth$; \citealt{Mills2019}), are the highest-S/N detections in the sample but lie outside its sub-Neptune focus. The remaining upper limits are consistent with the literature: Kepler-65\,b--d with the photodynamical masses of \citet{Mills2019}, Kepler-126\,b and c and Kepler-129\,b with the HIRES minimum masses of \citet{Weiss2023}, and Kepler-145\,c with the TTV mass of \citet{Ofir2025}. Kepler-129\,c is bounded below $39\,\mearth$ at 95\%, against the HIRES minimum mass of $25 \pm 9\,\mearth$ \citep{Weiss2023}. For TOI-1411\,b the 95\% limit of $2.0\,\mearth$ equals the median HIRES RV mass of \citet{Polanski2024}.

The most consequential comparison is Kepler-126\,d. Its published minimum mass of $55 \pm 23\,\mearth$, from 58 Keck/HIRES epochs ($K = 7.0 \pm 3.0\,\mathrm{m\,s^{-1}}$; \citealt{Weiss2023}), corresponds to a bulk density of the order of $20\,\mathrm{g\,cm^{-3}}$ at this radius. The stacked RV posterior instead gives $K = 0.50^{+0.93}_{-0.33}\,\mathrm{m\,s^{-1}}$, an upper limit that holds the mass below $17\,\mearth$ at 95\% confidence and disfavours the published value at the $2.1\sigma$ level. The RV-preferred density is $1.42^{+2.75}_{-0.94}\,\mathrm{g\,cm^{-3}}$, inconsistent with the $\sim\!20\,\mathrm{g\,cm^{-3}}$ implied by the published mass, thus ruling out the extreme-density regime. Kepler-145\,b is the only other offset of this size: its TTV mass of $34^{+25}_{-11}\,\mearth$ \citep{Ofir2025} differs from our RV mass posterior ($M_p<21\,\mearth$ at 95\%) at the $2.0\sigma$ level.

With four transiting detections the sample cannot support a density-distribution or bimodality test, such as a comparison with the simulations of \citet{Burn2024b}, and we defer it to a future study. This work contributes new constraints on 12 systems and a methodological precedent that can be applied to larger samples.

\subsection{Non-transiting companion candidates}
\label{sec:free_planet_results}

No additional planet is found in any of the 12 systems: all 24 free-planet slots evaluated (two per target) are classified as spurious. The closest case is K2-66, with $\ln B_{1,\,0} \geq 4.3$ (a lower bound, as its $N_\mathrm{p}=0$ bin is empty), short of the threshold of 5. The signal converges on a MAP period of 1503\,d (median 1516\,d, 68\% CI 1445--1626\,d) with $K/\sigma_K = 3.7$, but fails the period-unimodality and skewness sub-criteria of Tier~2 (0.65 against the required 0.9, and 1.9 against the required 1). A longer baseline is needed to establish whether the K2-66 trend is astrophysical.
The absence of new companions is consistent with massive planets inside 1\,au being rare around hosts of inner super-Earths \citep{Bryan2019,Zhu2018}.

\section{Summary and conclusions} \label{sec:conclusion}

We present a systematic mass-measurement campaign for 12 sub-Neptune systems: K2-182, K2-184, K2-244, K2-269, K2-66, Kepler-126, Kepler-129, Kepler-145, Kepler-21, Kepler-538, Kepler-65, and TOI-1411. By applying a unified hierarchical Bayesian framework with homogeneous data preparation across all targets and five high-resolution spectrographs (HIRES, HARPS-N, ESPRESSO, APF, and NEID), we sought to produce a homogeneous sample of sub-Neptune masses. This type of sample is needed for population-level studies of composition and formation pathways.

Across the sample, we detected the reflex motion of four transiting planets, Kepler-21\,b, K2-66\,b, Kepler-538\,b, and K2-182\,b, together with the non-transiting giants Kepler-65\,e and Kepler-129\,d. The remaining fifteen signals yield upper limits, among them the first RV constraints on K2-244 and K2-269. Where prior masses exist, our values agree to within $2.1\sigma$, the largest offsets being Kepler-126\,d ($<\!17$ against $55 \pm 23\,\mearth$) and Kepler-145\,b ($<\!21$ against a TTV mass of $34^{+25}_{-11}\,\mearth$). Kepler-21\,b is consistent with an Earth-like rocky composition, whereas K2-66\,b, K2-182\,b, and Kepler-538\,b require a volatile component. No new non-transiting companion is found in any of the 12 systems, consistent with massive planets on short-to-intermediate periods not being the dominant architecture among these sub-Neptune systems.

The methodological contribution stands on its own. Combining four RV reductions of the same spectra through LOO-CV stacking yields masses that do not depend on a single pipeline choice. With respect to spectra that all four pipelines are able to reduce, the stacked semi-amplitude never excludes a single-pipeline result, while a reduction that gives a poor fit to a specific star receives a negligible weight without intervention. These weights depend on which RVs enter the fit, so they do not rank the pipelines. Such a ranking needs a multi-reduction comparison on a larger sample of high-S/N spectra. For demographic work, masses that do not depend on the choice of pipeline matter more than extra precision on any one target. The present sample is too small and unconstrained to be sufficient for testing the density distribution of FGK sub-Neptunes, but this work does represent the first step towards studies on larger samples of stars such as those of the THIRSTEE project \citep{Lacedelli2024}. This methodology will be used in subsequent works to provide mass constraints of an emerging census of sub-Neptunes capable of testing formation theories with homogeneous bulk density distributions.

\section*{Data availability}
The RVs of all four reductions (CCF, \texttt{SERVAL}, \texttt{LBL}, and \texttt{sBART}) for every target, with their quality flags, the HIRES, APF, and NEID RVs, and the activity indicators used in Sect.~\ref{stellar_rotation} are only available in electronic form at the Centre de Donn\'ees astronomiques de Strasbourg (CDS) via anonymous ftp to \url{cdsarc.u-strasbg.fr} (130.79.128.5) or via \url{http://cdsweb.u-strasbg.fr/cgi-bin/qcat?J/A+A/}.

\begin{acknowledgements}

Some of the data presented herein were obtained at Keck Observatory, which is a private 501(c)3 non-profit organization operated as a scientific partnership among the California Institute of Technology, the University of California, and the National Aeronautics and Space Administration. The Observatory was made possible by the generous financial support of the W. M. Keck Foundation. The raw Keck Observatory data presented herein, once public, can be accessed from the Keck Observatory Archive (KOA), which is operated by the W. M. Keck Observatory and the NASA Exoplanet Science Institute (NExScI), under contract with the National Aeronautics and Space Administration. The authors wish to recognize and acknowledge the very significant cultural role and reverence that the summit of Maunakea has always had within the Native Hawaiian community. We are most fortunate to have the opportunity to conduct observations from this mountain.

Data presented herein were obtained with the WIYN 3.5m from telescope time allocated under the NN-EXPLORE program, with support from the National Aeronautics and Space Administration and the U.S. National Science Foundation. Data presented here were obtained by the NEID spectrograph built by Penn State University and operated at the WIYN Observatory by NSF NOIRLab, under the NN-EXPLORE partnership of the National Aeronautics and Space Administration and the U.S. National Science Foundation. The authors are honored to be permitted to conduct astronomical research on I'oligam Du’ag (Kitt Peak), a mountain with particular significance to the Tohono O’odham.

P.L., R.L., and P.F. acknowledge financial support from the Severo Ochoa grant CEX2021-001131-S funded by MCIN/AEI/10.13039/501100011033 and by the European Union (ERC Starting Grant THIRSTEE, 101164189). 

We also acknowledge financial support from the Agencia Estatal de Investigaci\'on of the Ministerio de Ciencia e Innovaci\'on MCIN/AEI/10.13039/501100011033 and the ERDF “A way of making Europe” through projects PID2021-125627OB-C32 and PID2024-158486OB-C32. This work is supported by the ERC Grant (ERC Advanced Grant SPEAR, GA 101200674). Funded by the European Union. Views and opinions expressed are however those of the author(s) only and do not necessarily reflect those of the European Union or the  European Research Council Executive Agency (ERCEA). Neither the European Union nor the granting authority can be held responsible for them.

We used the following \texttt{Python} packages for data analysis and visualisation: \texttt{numpy} \citep{Harris2020}, \texttt{scipy} \citep{Virtanen2020}, \texttt{matplotlib} \citep{Hunter2007}, \texttt{astropy} \citep{TheAstropyCollaboration2022}, \texttt{UltraNest} \citep{Buchner2021}, and \texttt{kima} \citep{Faria2018}.

\end{acknowledgements}

\bibliographystyle{aa}
\bibliography{bibliography}

\begin{appendix}
\onecolumn

\section{Homogeneity of the adopted stellar parameters}\label{app:ariadne}

\begin{figure}[H]
\centering
\includegraphics[width=0.78\linewidth]{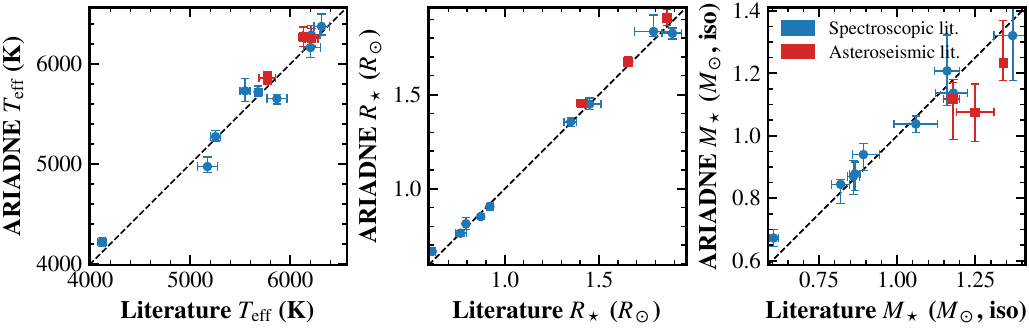}
\caption{ARIADNE SED-fit stellar parameters versus adopted literature values for our sample ($T_\mathrm{eff}$, $R_\star$, $M_\star$). The asteroseismic targets are Kepler-21, Kepler-65, and Kepler-129. ARIADNE results are consistent with the literature within $1$--$2\sigma$ across most of the sample. The largest offsets occur for asteroseismically constrained targets in $R_\star$ and $M_\star$, where the precision of SED fitting is surpassed.}
\label{fig:ariadne_comparison}
\end{figure}

\section{Planetary parameters}\label{app:params}

\begin{table}[H]
\caption{Planetary parameters for our sample.}
\label{tab:masses}
\centering
\small
\setlength{\tabcolsep}{3pt}
\resizebox{\textwidth}{!}{%
\begin{tabular}{l l c c c c c c c c l l}
\hline\hline
Star & Planet & $P$ (d) & $K^{\rm lit}$ (m\,s$^{-1}$) & $K$ (m\,s$^{-1}$) & $e$ & $R_p^{\rm lit}$ ($\rearth$) & $M_p^{\rm lit}$ ($\mearth$) & $M_p$ ($\mearth$) & $\rho_p$ (g\,cm$^{-3}$) & $R$ ref. & $M$ ref. \\
\hline
K2-182 & b & $4.7369696 \pm 0.0000017$ & $9.1 \pm 2.3$ & $8.47^{+1.85}_{-3.22}$ & $0.05^{+0.11}_{-0.04}$ & $2.744 \pm 0.065$ & $21.0 \pm 5.4$ & $20.1^{+4.5}_{-7.8}$ & $5.33^{+1.29}_{-2.06}$ & Thy23 & Thy23 \\
\hline
K2-184 & b & $16.978010 \pm 0.000024$ & $4.4^{+1.5}_{-1.6}$ & $<\!2.7$ & $0.18^{+0.19}_{-0.13}$ & $1.235 \pm 0.077$ & $10.9 \pm 4.7$ & $<\!9.1$ & $<\!29$ & Hob26 & Hob26 \\
\hline
K2-244 & b & $21.0688 \pm 0.0032$ & \ldots & $<\!2.8$ & $0.18^{+0.29}_{-0.14}$ & $1.79 \pm 0.12$ & \ldots & $<\!9.6$ & $<\!10$ & Liv18 &  \\
\hline
K2-269 & b & $4.14497^{+0.00082}_{-0.00062}$ & \ldots & $<\!2.5$ & $0.18^{+0.28}_{-0.14}$ & $1.59 \pm 0.13$ & \ldots & $<\!5.9$ & $<\!8.6$ & Liv18b &  \\
\hline
K2-66 & b & $5.06963 \pm 0.00081$ & $5.9 \pm 1.3$ & $6.48 \pm 1.17$ & $0.10^{+0.12}_{-0.07}$ & $2.75 \pm 0.16$ & $16.0 \pm 4.0$ & $17.9 \pm 3.2$ & $4.69 \pm 1.31$ & How25 & How25 \\
\hline
Kepler-21 & b & $2.785780 \pm 0.000030$ & $2.41 \pm 0.49$ & $2.13^{+0.25}_{-0.35}$ & $0.07^{+0.10}_{-0.05}$ & $1.600 \pm 0.020$ & $6.7 \pm 1.4$ & $5.66^{+0.65}_{-0.96}$ & $7.61^{+0.96}_{-1.28}$ & LM16 & BR24 \\
\hline
\multirow{4}{*}{Kepler-65} & b & $2.1549209 \pm 0.0000086$ & $1.0 \pm 1.0$ & $<\!3.2$ & $0.14^{+0.23}_{-0.11}$ & $1.419 \pm 0.031$ & $2.4^{+2.4}_{-1.6}$ & $<\!7.1$ & $<\!14$ & Mil19 & Mil19 \\
 & c & $5.859697 \pm 0.000093$ & $1.5 \pm 1.2$ & $<\!2.6$ & $0.14^{+0.20}_{-0.10}$ & $2.576 \pm 0.056$ & $5.4 \pm 1.7$ & $<\!8.1$ & $<\!2.7$ & &  \\
 & d & $8.13167 \pm 0.00024$ & $1.2 \pm 1.1$ & $<\!2.3$ & $0.27^{+0.25}_{-0.21}$ & $1.557 \pm 0.059$ & $4.14 \pm 0.79$ & $<\!7.5$ & $<\!11$ & &  \\
 & e & $258.8 \pm 1.5$ & $19.1 \pm 1.8$ & $18.85^{+1.11}_{-1.45}$ & $0.24^{+0.05}_{-0.12}$ & \ldots & $207.5^{+17.8}_{-17.5}$ & $212^{+11}_{-14}$\tablefootmark{a} & \ldots & &  \\
\hline
\multirow{3}{*}{Kepler-126} & b & $10.495678 \pm 0.000017$ & $0.1 \pm 1.9$ & $<\!2.4$ & $0.12^{+0.23}_{-0.10}$ & $1.460^{+0.057}_{-0.071}$ & $4.1 \pm 7.1$ & $<\!8.9$ & $<\!16$ & Mor16 & Wei24 \\
 & c & $21.869676 \pm 0.000054$ & $0.3 \pm 1.5$ & $<\!1.5$ & $0.16^{+0.26}_{-0.13}$ & $1.56 \pm 0.11$ & $1.3 \pm 7.0$ & $<\!7.1$ & $<\!11$ & &  \\
 & d & $100.28287 \pm 0.00017$ & $7.0 \pm 3.0$ & $<\!2.2$ & $0.16^{+0.24}_{-0.12}$ & $2.461 \pm 0.064$ & $55 \pm 23$ & $<\!17$ & $<\!6.4$ & &  \\
\hline
\multirow{3}{*}{Kepler-129} & b & $15.791937 \pm 0.000036$ & $2.3 \pm 1.4$ & $<\!3.6$ & $0.14^{+0.23}_{-0.12}$ & $2.321 \pm 0.046$ & $10.2 \pm 6.2$ & $<\!15$ & $<\!6.8$ & Mor16 & Wei24 \\
 & c\tablefootmark{c} & $82.20001 \pm 0.00043$ & $3.2 \pm 1.2$ & $<\!5.4$ & $0.17^{+0.18}_{-0.13}$ & $2.449^{+0.073}_{-0.055}$ & $25.3 \pm 9.3$ & $<\!39$ & $<\!15$ & &  \\
 & d & $1936 \pm 25$ & $90.0 \pm 1.9$ & $91.62^{+1.30}_{-1.70}$ & $0.07 \pm 0.02$ & \ldots & $2015 \pm 43$ & $1986^{+29}_{-35}$\tablefootmark{a} & \ldots & &  \\
\hline
\multirow{2}{*}{Kepler-145} & b & $22.950623 \pm 0.000099$ & \ldots & $<\!4.1$ & $0.17^{+0.25}_{-0.13}$ & $2.32 \pm 0.23$ & $34^{+25}_{-11}$ & $<\!21$ & $<\!10$ & Ofi25 & Ofi25 \\
 & c & $42.882480 \pm 0.000081$ & \ldots & $<\!11$ & $0.14^{+0.20}_{-0.10}$ & $3.86 \pm 0.23$ & $25.1^{+9.4}_{-6.0}$ & $<\!73$ & $<\!7.6$ & &  \\
\hline
\multirow{2}{*}{Kepler-538} & b & $81.73778 \pm 0.00013$ & $2.09 \pm 0.47$ & $1.74 \pm 0.39$ & $0.11^{+0.14}_{-0.08}$ & $2.219 \pm 0.038$ & $12.9 \pm 2.9$ & $10.8^{+2.4}_{-2.8}$ & $5.47 \pm 1.22$ & May19 & Bon23 \\
 & c\tablefootmark{b} & $117.7610 \pm 0.0022$ & \ldots & $<\!0.74$ & $0.15^{+0.27}_{-0.12}$ & $0.616^{+0.053}_{-0.065}$ & \ldots & $<\!5.0$ & $<\!134$ & FP18 &  \\
\hline
TOI-1411 & b & $1.4520527 \pm 0.0000017$ & $1.55 \pm 0.93$ & $<\!1.7$ & $0.15^{+0.27}_{-0.12}$ & $1.203 \pm 0.048$ & $2.0 \pm 1.2$ & $<\!2.0$ & $<\!6.4$ & Pol24 & Pol24 \\
\hline
\end{tabular}%
}
\tablefoot{Reference abbreviations: Thy23 = \citet{Thygesen2023}; Hob26 = \citet{Hobson2026}; Liv18 = \citet{Livingston2018}; Liv18b = \citet{Livingston2018b}; How25 = \citet{Howard2025}; LM16 = \citet{LopezMorales2016}; Mil19 = \citet{Mills2019}; Mor16 = \citet{Morton2016}; Ofi25 = \citet{Ofir2025}; May19 = \citet{Mayo2019}; FP18 = \citet{Fulton2018}; Pol24 = \citet{Polanski2024}; BR24 = \citet{BeardRobertson2024}; Wei24 = \citet{Weiss2023}; Bon23 = \citet{Bonomo2023}. $R_p^{\rm lit}$ is the published $R_p/R_\star$ (source in the $R$ ref. column; the radius itself for K2-66\,b) multiplied by our adopted $R_\star$ (Table~\ref{tab:sample}). $K^{\rm lit}$ and $M_p^{\rm lit}$ come from the paper in the $M$ ref. column, given once per system, except $K^{\rm lit}$ of Kepler-65\,b--d (Wei24). $K$, $M_p$ and $\rho_p$ are 95\% upper limits ($<$) where $K/\sigma_K\leq3$. \tablefootmark{a}Non-transiting, minimum mass $M_p\sin i$. \tablefootmark{b}Unconfirmed candidate (KOI-365.02). \tablefootmark{c}$K/\sigma_K = 3.0$, just under the detection threshold.}
\end{table}

\section{Noise-model selection and GP hyperparameters}\label{app:noise}

\begin{table}[H]
\centering
\caption{Bayesian model-comparison grid for the RV noise model.}
\label{tab:dlnz}
\footnotesize
\begin{tabular}{l c c c c c c c}
\hline\hline
Target & $N_{\rm pl}$ & poly0 & poly1 & poly2 & $+$GP poly0 & $+$GP poly1 & $+$GP poly2 \\
\hline
K2-182 & 1 & 0.0$^\dagger$ & -5.6 & -7.6 & -0.1 & -5.7 & -8.9 \\
K2-184 & 1 & -2.6$^\dagger$ & -2.9 & -3.9 & 0.0 & -0.4 & -1.2 \\
K2-244 & 1 & 0.0$^\dagger$ & -1.4 & -1.8 & -0.1 & -1.4 & -1.8 \\
K2-269 & 1 & -0.1$^\dagger$ & -1.5 & -2.9 & 0.0 & -1.5 & -2.9 \\
K2-66 & 1 & -45.0 & -7.0 & -7.9 & -0.3$^\dagger$ & 0.0 & -0.2 \\
Kepler-126 & 3 & 0.0$^\dagger$ & -8.3 & -15.9 & -0.1 & -7.9 & -13.1 \\
Kepler-129 & 3 & 0.0$^\dagger$ & -3.4 & -1.5 & -1.3 & -1.3 & -2.3 \\
Kepler-145 & 2 & 0.0$^\dagger$ & -1.5 & -0.7 & 0.0 & -1.3 & -0.8 \\
Kepler-21 & 1 & -105.5 & -98.3 & -172.1 & -11.2 & 0.0$^\dagger$ & -36.6 \\
Kepler-538 & 2 & -3.9$^\dagger$ & -13.2 & -34.0 & 0.0 & -8.6 & -13.6 \\
Kepler-65 & 4 & 0.0$^\dagger$ & -14.8 & -49.0 & -0.7 & -6.3 & -9.0 \\
TOI-1411 & 1 & -1.0$^\dagger$ & -9.4 & -16.5 & 0.0 & -6.8 & -14.3 \\
\hline
\end{tabular}
\tablefoot{$\Delta\ln\mathcal{Z}$ is the mean over reductions for multi-reduction targets, given relative to each target's highest-evidence model. $^\dagger$ marks the model adopted under the parsimony rule of Sect.~\ref{sec:modelling}.}
\end{table}

\begin{table}[H]
\centering
\caption{Posteriors of the quasi-periodic GP hyperparameters for Kepler-21 and K2-66.}
\label{tab:gp_hyper}
\begin{tabular}{l c c c c}
\hline\hline
 & $\eta_1$ (amplitude) & $\eta_2$ (timescale) & $\eta_3$ (rotation) & $\eta_4$ (smoothing) \\
Target & [m\,s$^{-1}$] & [d] & [d] & \\
\hline
K2-66 & $27.8^{+11.2}_{-7.7}$ & $303^{+116}_{-110}$ & $27.75^{+1.16}_{-1.08}$ & $6.63^{+2.24}_{-2.46}$ \\
Kepler-21 & $4.2^{+0.6}_{-0.5}$ & $42^{+436}_{-24}$ & $13.50^{+0.03}_{-0.60}$ & $0.63^{+0.19}_{-0.12}$ \\
\hline
\end{tabular}
\end{table}

\section{RV pipeline comparison}\label{app:pipelines}

\begin{figure}[H]
\centering
\includegraphics[width=\linewidth]{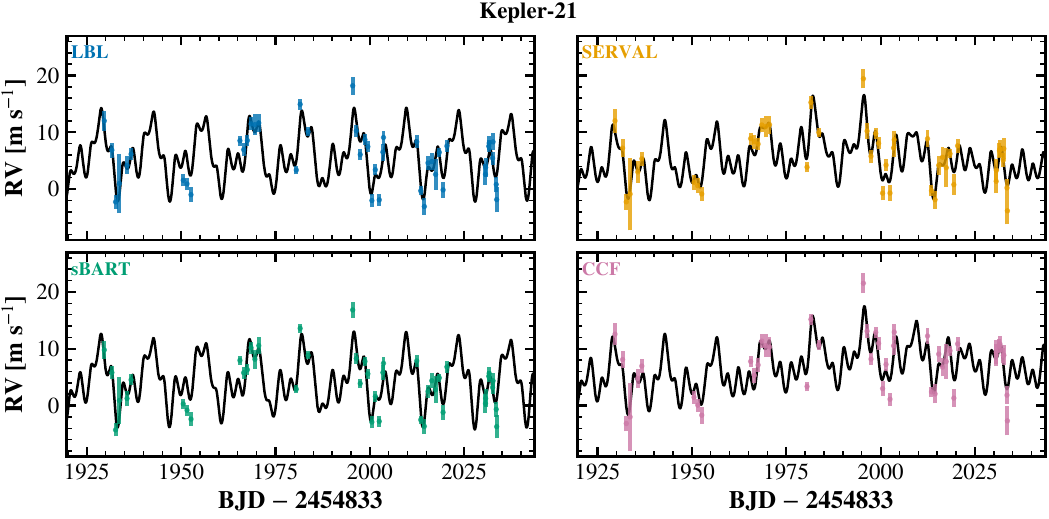}
\caption{Per-pipeline HARPS-N RV time series for Kepler-21 over the densest 130\,d window, each with its own model (Keplerian + GP + linear trend). The quasi-periodic activity signal is recovered in all reductions. The stacking weights of the fit including the HIRES, APF, and NEID RVs are $w_\mathrm{sBART} = 58\%$, $w_\mathrm{SERVAL} = 42\%$, $w_\mathrm{LBL} = 0\%$ and $w_\mathrm{CCF} = 0\%$.}
\label{fig:kepler21_rv_comparison}
\end{figure}

\begin{figure}[H]
\centering
\includegraphics[width=0.90\linewidth]{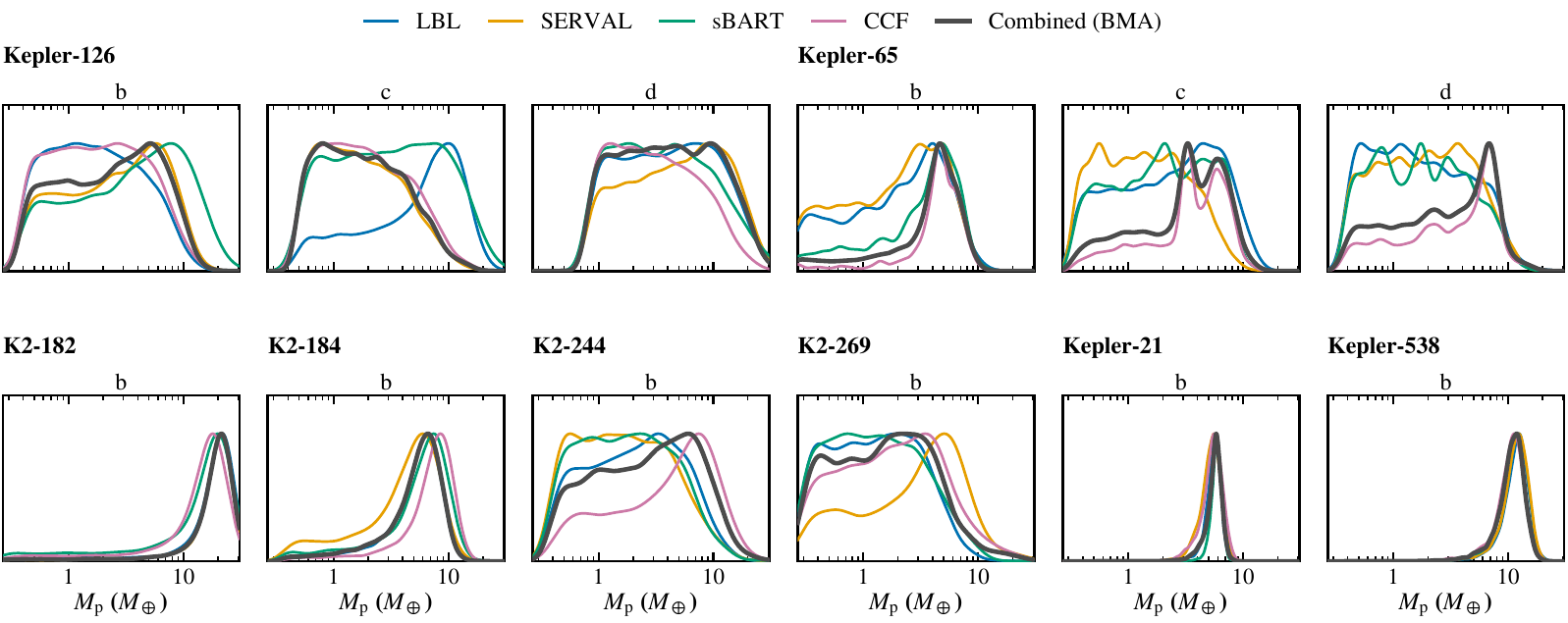}
\caption{Mass posteriors of the sub-Neptunes in the eight multi-reduction systems for each reduction (full dataset), with the stacked posterior in grey. The reductions agree for the detected planets (Kepler-21\,b, Kepler-538\,b, K2-182\,b) and differ mainly in the unconstrained tails of the non-detections.}
\label{fig:mosaic_reduction}
\end{figure}

Figure~\ref{fig:kepler21_rv_comparison} compares the per-pipeline time series of Kepler-21, the worked example of the multi-pipeline framework (full RV model in Fig.~\ref{fig:kepler21_rv_hero}). Figure~\ref{fig:mosaic_reduction} shows the mass posteriors of the four reductions for the sub-Neptunes of all multi-reduction targets.

\end{appendix}
\end{document}